\documentclass[a4paper,twocolumn,10pt,accepted=2023-10-25]{quantumarticle}
\pdfoutput=1

\usepackage[sort&compress,numbers,square,comma]{natbib}

\usepackage{textgreek}
\usepackage[T1]{fontenc}
\usepackage[utf8]{inputenc}
\usepackage[english]{babel}
\usepackage{microtype}
\usepackage{indentfirst}	
\usepackage{amsthm, amssymb, amstext, amsmath, bbm}
\usepackage{dsfont}
\usepackage{cancel}
\usepackage{float}
\usepackage{tikz}
\usepackage{soul}
\usepackage{epstopdf}
\usepackage{blkarray} 
\usepackage{booktabs}
\usepackage{braket}
\usepackage{xcolor}
\usepackage{array}

\definecolor{darkred}{rgb}{0.5,0,0}
\definecolor{darkgreen}{rgb}{0,0.5,0}
\definecolor{darkblue}{rgb}{0,0,0.5}

\usepackage[colorlinks]{hyperref}
\hypersetup{colorlinks,
linkcolor=darkred,
filecolor=darkgreen,
urlcolor=darkblue,
citecolor=darkgreen}

\newcommand{\LL}{\mathcal{L}}
\newcommand{\DD}{\mathcal{D}}

\newcommand{\rhot}{\hat{\rho}(t)}

\newcommand{\expec}[1]{\langle #1 \rangle}

\renewcommand{\Re}[1]{\mathbb{R}\mathrm{e}\left[#1\right]}

\newcommand{\sss}{\hat{\rho}_{\rm ss}}
\newcommand{\eig}[1]{\hat{\rho}_{#1}}
\newcommand{\symmat}[1]{\hat{\tilde{\rho}}_{#1}}

\newcommand{\abs}[1]{\lvert #1 \rvert}

\newcommand{\Tr}[1]{\mathrm{Tr}\!\left[#1\right]}
\usepackage{epstopdf}

\makeatletter
\newcommand*\bigcdot{\mathpalette\bigcdot@{.5}}
\newcommand*\bigcdot@[2]{\mathbin{\vcenter{\hbox{\scalebox{#2}{$\m@th#1\bullet$}}}}}
\makeatother

\usepackage{bbm}

\renewcommand{\eqref}[1]{Eq.~(\ref{#1})}

\usepackage{mathtools}

\DeclarePairedDelimiter\floor{\lfloor}{\rfloor}

\begin{document}
\title{Dissipative phase transitions in $n$-photon driven quantum nonlinear resonators}

\author{Fabrizio Minganti}\email{fabrizio.minganti@epfl.ch }
\affiliation{Institute of Physics, Ecole Polytechnique F\'ed\'erale de Lausanne (EPFL), CH-1015 Lausanne, Switzerland}
\affiliation{Center for Quantum Science and Engineering, Ecole Polytechnique F\'ed\'erale de Lausanne (EPFL), CH-1015 Lausanne, Switzerland}

\author{Vincenzo Savona}\email{vincenzo.savona@epfl.ch }
\affiliation{Institute of Physics, Ecole Polytechnique F\'ed\'erale de Lausanne (EPFL), CH-1015 Lausanne, Switzerland}
\affiliation{Center for Quantum Science and Engineering, Ecole Polytechnique F\'ed\'erale de Lausanne (EPFL), CH-1015 Lausanne, Switzerland}

\author{Alberto Biella}
\email{alberto.biella@ino.cnr.it}
\affiliation{Pitaevskii BEC Center, CNR-INO and Dipartimento di Fisica, Università di Trento, I-38123 Trento, Italy}

\date{2023/10/31}

\begin{abstract}
We investigate and characterize the emergence of finite-component dissipative phase transitions (DPTs) in nonlinear photon resonators subject to $n$-photon driving and dissipation.
Exploiting a semiclassical approach, we derive general results on the occurrence of second-order DPTs in this class of systems.
We show that for all odd $n$, no second-order DPT can occur while, for even $n$, the competition between higher-order nonlinearities determines the nature of the criticality and allows for second-order DPTs to emerge only for $n=2$ and $n=4$. 
As pivotal examples, we study the full quantum dynamics of three- and four-photon driven-dissipative Kerr resonators, confirming the prediction of the semiclassical analysis on the nature of the transitions. 
The stability of the vacuum and the typical timescales needed to access the different phases are also discussed.
We also show a first-order DPT where multiple solutions emerge around zero, low, and high-photon numbers.
Our results highlight the crucial role played by {\it \textbf{strong}} and {\it \textbf{weak}} symmetries in triggering critical behaviors, providing a Liouvillian framework to study the effects of high-order nonlinear processes in driven-dissipative systems, that can be applied to problems in quantum sensing and information processing. 
\end{abstract}

\maketitle


\section{Introduction, motivations, and summary of the main results}

Nonlinear bosonic systems, such as optical cavities, polaritonic systems, optomechanical resonators, and superconducting circuits, represent an extremely rich and versatile tool to explore and simulate nonequilibrium quantum physics \cite{Carusotto_RMP_2013_quantum_fluids_light,Carusotto2020,LeHurCRP16}.
These systems are intrinsically {\it open}, meaning that particle, energy, and correlations can be gained or lost through the coupling with the environment \cite{BreuerBookOpen}.
Drives are then applied to these systems, bringing them out of their thermal equilibrium and compensating for the losses induced by the environment.
As a result, the complex interplay between driving, dissipation, and Hamiltonian terms dictates the system's dynamics and determines its steady states, whose properties can differ from those of closed quantum systems at equilibrium \cite{VerstraeteNATPH2009,DiehlNATPH2008,DiehlPRL10}.

The symmetries of the drive and dissipators play a fundamental role in determining both the nature of the steady state and the dynamical properties  of a quantum system \cite{BucaNPJ2012,AlbertPRA14,MingantPRA18_Spectral}.
For instance, in a nonlinear photonic cavity the possibility to exploit nonlinear and engineered pumping schemes, in the presence of moderate single-particle dissipation, opened venues to the generation and stabilization of nonclassical states \cite{BartoloPRA16,LebreuillyPRA17,BiellaPRA17}.
A pivotal example in this field is the use of two-photon drives to generate, stabilize, and control photonic Schr\"odinger cat states \cite{LeghtasScience15,grimm_stabilization_2020},
that have been proposed as a fundamental building block of quantum computing devices \cite{MirrahimiNJP14}.
Beyond their interest in quantum information, parametric processes have been at the center of intense research, leading to the exploration of their properties both in classical \cite{ChenPRL08,LeuchPRL16} and quantum configurations \cite{BartoloPRA16,BartoloEPJST17,GotoPRA16,Labay-MoraPRL22}.

The study and characterization of  dissipative phase transitions (DPTs) and their peculiarities have been the focus of a vast theoretical and experimental research, especially concerning the connection of DPTs to multimodality and metastability \cite{LandaPRL20}.
In this scenario, two main distinctions have been drawn in characterizing DPTs \cite{MingantPRA18_Spectral}.
First-order DPTs are discontinuous changes in the properties of the system's steady state as a function of a control parameter \cite{KesslerPRA12}. These have been associated with hysteresis and critical slowing down \cite{CasteelsPRA16,BartoloPRA16,RodriguezPRL17}, which allow one to observe the emergence of metastable dynamics and to study its competition with the other typical timescales of the system.
Key to understanding second-order DPTs -- where the steady state transitions continuously, but it is characterized by a divergent response function --  are symmetries \cite{BartoloPRA16,SavonaPRA17,RotaPRL19}.
In particular, DPTs can be associated with weak and strong symmetry breaking \cite{PRLLieu20,HalatiPRR22}.
Second-order DPTs are useful for several technological tasks.
The cross-fertilization between quantum information processing and open system criticality led to innovative ideas to protect quantum information \cite{PRLLieu20,gravina2022critical}, enhance quantum sensing \cite{Fernandez-Lorenzo2017QuantumResources, Ilias2022Criticality-EnhancedMeasurement, Raghunandan2018High-DensityTransitions,di2021critical}, and review laser theory \cite{Takemura2021,minganti2021liouvillian,Yacomotti2022}.

In the panorama of DPTs, the parametrically-driven (or two-photon) Kerr resonator has attracted considerable interest \cite{BartoloPRA16,MingantPRA18_Spectral,SavonaPRA17,RotaPRL19,HeugelPRL19}.
Indeed, it provided an ideal test model that displays both first- and second-order criticalities in different regions of the parameter space \cite{BartoloPRA16} and represents one of the few cases for which a steady-state can be analytically found \cite{MingantiSciRep16,RobertsPRX20}.
Thus, DPTs of this model were extensively studied  \cite{BartoloPRA16,BartoloEPJST17,SavonaPRA17,HeugelPRL19}, in connection with the spectral properties of the Liouvillian \cite{MingantPRA18_Spectral} and more exotic phenomena, such as exceptional points and parity-time symmetry breaking \cite{ZhangPRA21}.
These findings represented the natural extension of the well-known results about first-order DPTs in the coherently-driven (one-photon) Kerr resonator \cite{CasteelsPRA16,BartoloPRA16,RodriguezPRL17,FitzpatrickPRX17,FinkNatPhys18,Brookes2021,Chen22}, pioneered by Drummond and Walls \cite{Drummond_JPA_80_bistability}, and showed how the presence of multi-photon driving and dissipation can drastically modify the physics of nonlinear bosonic resonators \cite{BartoloPRA16,SavonaPRA17,PRLLieu20}.
Remarkably, all these results obtained at the single-resonator level provided a guideline to investigate emergent phenomena in more complex lattice architectures \cite{RotaPRL19,VicentiniPRA18,Foss-FeigPRA17,VerstraelenPRR20,RotaPRA19}.
The fundamental theoretical interest in studying the properties of single or few resonators \cite{CasteelsPRA16,CasteelsPRA17,CasteelsPRA17-2,MingantPRA18_Spectral,SavonaPRA17,MingantiPRA23,felicetti2020universal}, but with $n$-photon driving schemes, is also strongly motivated by the recent achievement of higher-order photon pumping exploiting strong nonlinearities in superconducting circuits \cite{Svensson2018,ChangPRX20,Lang2021}.

\subsection{Summary of the main results}

In this work, we advance these ideas and explore the DPTs of nonlinear photonic resonators in the presence of parametric $n$-photon drive and losses, going beyond the aforementioned $n=1,2$ cases.
We provide a general criterion for arbitrary $n$ to argue the nature of criticalities for this ubiquitous class of models in quantum optics.
To this aim, we prove a no-go theorem in the semiclassical limit, predicting the emergence and the nature of DPTs in this class of systems. 
We derive a general recipe to connect the order of the DPTs and the presence of certain Hamiltonian terms, highlighting the importance that strong and weak symmetry has in constraining the system dynamics, and we point out the technical limitations in witnessing second-order DPTs for $n>4$ driving schemes. 

We then test the semiclassical predictions
by performing a detailed numerical analysis of the full quantum model for the $n=3$ and $n=4$ cases.
We confirm the role of symmetries predicted by the semiclassical theory, and we analyze the DPTs within the theoretical framework of the spectral properties of the Liouvillian superoperator.
For $n=3$, we confirm the semiclassical prediction and show that the system can only undergo a first-order phase transition accompanied by the symmetry breaking of the discrete weak ${Z}_{3}$ symmetry, as the system parameters are scaled towards the thermodynamic limit.
For the 4-photon driven resonator, we show that both a first- and second-order DPT can occur, accompanied by a breaking of the ${Z}_{4}$, showing that 9 states can be stabilized across a first-order DPT, making these systems possible candidates for associative memories \cite{Labay-MoraPRL22}.

The paper is structured as follows.
In Sec.~\ref{sec:model}, we introduce the model, the master equation governing the driven-dissipative dynamics, the symmetry properties of the problem, and their consequences on the Liouvillian spectrum.
In Sec.~\ref{sec:semiclassical}, we discuss the emergence of DPTs in the semiclassical limit, while in Secs.~\ref{sec:3ph_quantum_kerr} and \ref{sec:4ph_quantum_kerr} we study the full quantum dynamics for $n=3,4$ resonators, respectively.
Finally, in Sec.~\ref{sec:con}, we draw our conclusions and discuss some future perspectives.

\section{The model}
\label{sec:model}

We consider a bosonic $n$-driven nonlinear resonator whose Hamiltonian reads
\begin{equation}
\label{Eq:nph}
     \hat{H}_{n} =  \sum_{m=1}^{m_{\rm max}} \frac{U_m}{m} \left( \hat{a}^\dagger \right)^m \hat{a}^m + G_n \left[\hat{a}^n  + \left(\hat{a}^\dagger\right)^n \right],
\end{equation}
where $\hat{a}$ ($\hat{a}^\dagger$) is the bosonic annihilation (creation) operator.
The interaction strengths $U_m$ sets the scale of $m$-photon processes. For instance, $U_1$  characterizes the energy of one photon in the resonator (in the frame rotating at the pump frequency) and rescales the term $\hat{a}^\dagger \hat{a}$, $U_2$ is a standard Kerr interaction, and so on.
As detailed also in Appendix~\ref{app:impl}, for a $n$-photon drive we should consider at least processes up to the order $m_{\rm max}=\floor*{n/2+1}$, where $\floor*{A}$ indicates the integer part of the number $A$.
As we will see in the following, the high-order $U_m$s play a fundamental role in determining the nature of the DPTs and, for this reason, need to be included in a minimal model.
$G_n$, instead, represents the $n$-photon drive amplitude.

Given the dissipative nature of the system,
and within the Born and Markov approximations, the system's dynamics is ruled by a (Gorini-Kossakowski-Sudarshan) Lindblad master equation \cite{Lindblad1976, Gorini1976} reading
(hereafter we set $\hbar=1$)
\begin{equation}
\label{lin}
   \partial_{t}\hat{\rho}(t)= \LL[\hat{\rho}(t)]=- i [\hat{H}_{n}, \hat{\rho}(t)] + \gamma \DD[\hat{a}] + \eta_n \DD[\hat{a}^n],
\end{equation}
with $ \DD[\hat{O}] = \hat{O}\hat{\rho}(t) \hat{O}^{\dagger}-\{\hat{O}^{\dagger}\hat{O},\hat{\rho}(t)\}/2$.
The first term in \eqref{lin} rules the coherent (unitary) part of the dynamics, and follows from \eqref{Eq:nph}, upon an appropriate rescaling of $U_m$ due to the dressing of the cavity eigenmodes by the environment (Lamb-shift-like terms \cite{BreuerBookOpen}).
The second and the third terms in \eqref{lin} account for the incoherent one- and $n$-photon losses, respectively.
While one-photon dissipation is an unavoidable feature in any photonic resonator, emerging from the coupling of the cavity modes with the electromagnetic vacuum, $n$-photon losses naturally emerge as a byproduct of the engineered processes leading to $n$-photon drive.
Notice that other $m$-photon (with $m\neq1, \, n$) dissipative processes can be safely neglected, because their emergence is linked to the presence of engineered $m$-photon exchanges, and here we are considering only a single drive acting at each time.
Although the Hamiltonian in \eqref{Eq:nph} is quite general and platform-independent, we provide a brief discussion on how such terms can emerge in a superconducting circuit implementation in Appendix~\ref{app:impl}.

\subsection{Liouvillian spectrum, symmetries, and their breaking} 
\label{sec:symmetries}

Our analysis will mainly focus on the steady states $\sss^{(k)}$, i.e., that density matrices that do not evolve anymore under the action of the Lindblad master equation (\ref{lin}), defined by
\begin{equation}
    \partial_t \sss^{(k)} = \LL \sss^{(k)} =0.
\end{equation}
$k$ is an index that labels these steady states and in the present analysis will be solely tied to the presence of a strong symmetry (see below).
Otherwise, the steady state is unique and will be simply called $\sss$.

DPTs occur when the steady state of an open quantum system can display a nonanalytical behavior as a function of a generic parameter $\zeta$ \cite{KesslerPRA12,MingantPRA18_Spectral}.
As DPTs cannot occur in a finite-size system, one needs to investigate the so-called thermodynamic limit, formally defined as $L \to \infty$. 
While in a lattice system, one can think at $L$ as the number of sites, in the case under consideration scaling $L$ towards the thermodynamic limit implies a rescaling of the system parameters, as detailed in Sec.~\ref{subsec:thermodynamic}.
The non-analytical change of the steady-state is then  witnessed by the expectation value of some operator $\hat{o}$ as $\zeta$ crosses the critical point $\zeta_c$.
As such, we say that there is a phase transition of order $M$ if \cite{MingantPRA18_Spectral}
\begin{equation}\label{Eq:DPTDefinition}
\lim_{\zeta \to  \zeta_c}\left| \lim_{L \to \infty}  \frac{\partial^M}{\partial \zeta^M} \Tr{\sss^{(k)}(\zeta, L) \hat{o}}\right|=+ \infty.
\end{equation}

The $n$-photon driven Kerr resonator explicitly displays a ${Z}_n$ symmetry \footnote{We will use the notation ${Z}_n$ for the symmetry group, $\hat{Z}_n$ for the operator associated with such a symmetry, and $\mathcal{Z}_n$ for the corresponding superoperator.}. That is, the transformation 
\begin{equation}
\hat{a}\to \hat{a} \  e^{i 2 \pi k/n}, \quad k=0, \, 1,\, \dots , \,n,
\end{equation}
leaves the master equation (\ref{lin}) unchanged.
However, one can define two types of symmetries in open quantum systems \cite{BucaNPJ2012,AlbertPRA14}.
For the model under consideration, these are defined according to the way the operator $\hat{Z}_n =e^{i 2 \pi \hat{a}^\dagger \hat{a}/n}$ acts. One speaks of \textit{strong symmetries} if $\hat{Z}_n$ commutes with both the Hamiltonian and the jump operators, i.e.:
\begin{equation}\label{Eq:strong_symmetry}
[\hat{Z}_n, \hat{H}]=[\hat{Z}_n,  \hat{a}]=[\hat{Z}_n, \hat{a}^n]=0. 
\end{equation}
In this case, ${Z}_n$ implies the existence of a corresponding conserved quantity $\expec{\hat{Z}_n}_t\equiv{\rm Tr}[\hat{\rho}(t)\hat{Z}_n] = {\rm const}$. The system will display $n$ independent steady states, each one characterized by a different value of $\expec{\hat{Z}_n}_{\rm ss}\equiv \lim_{t\to\infty}\expec{\hat{Z}_n}_t$.
In our case, such a condition is fulfilled if and only if $\gamma=0$ (i.e., the photons are never lost individually).
The presence of a strong symmetry implies that there exist \textit{two superoperators} $\mathcal{Z}_n^{\rm L} = \hat{Z}_n \bigcdot \hat{\mathds{1}}$ and 
$\mathcal{Z}_n^{\rm R} = \hat{\mathds{1}}\bigcdot \hat{Z}_n$
\footnote{
The $\bigcdot$ notation for superoperators indicates that, if $\mathcal{S} = \hat{A} \bigcdot \hat{C}$, then
$\mathcal{S} \hat{B} = \hat{A} \hat{B} \hat{C}$. Details can be found in Ref.~\cite{Carmichael_BOOK_2}.
},
 such that 
\begin{equation}
\label{condstrong}
[\LL, \mathcal{Z}_n^{\rm L, \, R}]=0.
\end{equation}

A \textit{weak symmetry}, instead, does not respect the conditions in \eqref{Eq:strong_symmetry}, and as such the symmetry of the model does not entail a conserved quantity, meaning that $\expec{\hat{Z}_n}_t$ changes in time \cite{AlbertPRA14,BucaNPJ2012}. However, the superoperator $\mathcal{Z}_n =\hat{Z}_n  \bigcdot \hat{Z}_n $ commutes with the Liouvillian, i.e.
\begin{equation}
\label{condweak}
[\LL, \mathcal{Z}_n]=0.
\end{equation}

As a consequence of the conditions in Eqs.~(\ref{condstrong})~and~(\ref{condweak}), strong and weak symmetries constrain the structure of the Liouvillian $\LL$ and of its spectrum.
A compact and convenient way to discuss symmetries and phase transitions is via the spectral properties of the Liouvillian \cite{MingantPRA18_Spectral}. 
Given any Liouvillian $\LL$, we can introduce its eigenvalues $\lambda_i$ and right eigenoperators $\eig{i}$, defined via the relation
\begin{equation}\label{Eq:Spectrum}
\LL \eig{i}=\lambda_i \eig{i},
\end{equation} where $\Re{\lambda_i}\leq 0, \forall i$ represents the decay rates induced by the dissipative dynamics \cite{BreuerBookOpen,RivasBOOK_Open}.

\begin{figure*}
    \centering
    \includegraphics[width=0.98 \textwidth]{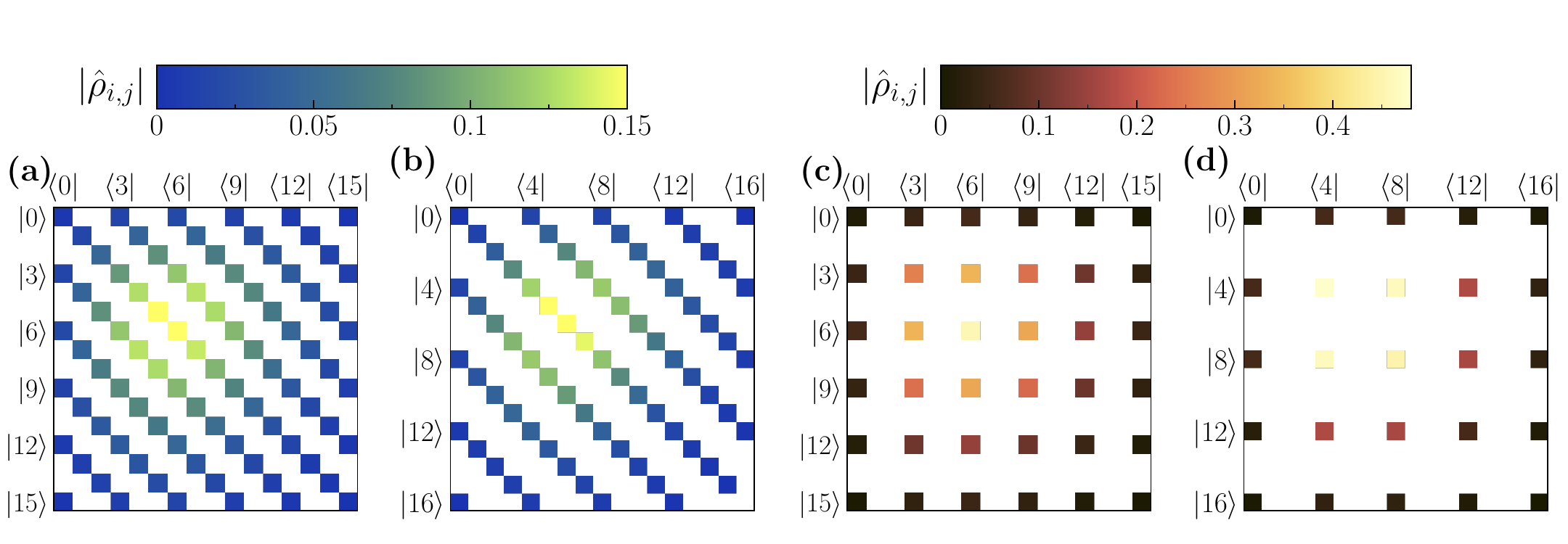}
    \caption{Sketch of the structure of (one of the) steady-state density matrix (matrices) $\eig{0}^{(0)}$ ($\eig{0}^{(0,0)}$) for a weak (a, b) and a strong (c, d) $Z_3$ (a,c) and $Z_4$ (b,d) symmetries. White indicates that the matrix element is zero. 
    Comparing the weak and strong symmetric cases, one notices the effect of one-photon dissipation: the breaking of the strong symmetry by $\gamma$ results in a  mixing of the populations, and thus in more nonzero elements. Nonetheless, being an incoherent process, no coherences between different symmetry sectors can be retained, thus resulting in a coarser steady-state structure.
    Hamiltonian parameters: (a, c) $U_1 = 5 \eta_3$, $U_2 = 3 \eta_3$, $G_3 = 9 \eta_3$; 
    (b, d) $U_1 = 5 \eta_4$, $U_2 = 3 \eta_4$, $U_3 = \eta_4/5$, $G_4 = 20 \eta_4$.
    Dissipation: (a) $\gamma = \eta_3$, (b) $\gamma = \eta_4$.
    }
    \label{fig:checkerboards}
\end{figure*}

\subsubsection{DPTs and weak symmetries}

The presence of a weak symmetry allows for refining the discussion on the spectral properties of the system.
The eigenvalues $z_n^{(k)}$ of $\mathcal{Z}_n$ are the $n$ roots of the unity [indeed, $(\mathcal{Z}_n)^{n}=1$], that is $z_n^{(k)}=e^{2 i \pi k /N}$ for $k=0, \, 1,\dots n-1$.
Since each eigenstate of $\LL$ must also be eigenstate of $\mathcal{Z}_n$, we can introduce the ``quantum number'' $k$, such that, for a weak $Z_n$ symmetry,
\begin{equation}
 \mathcal{Z}_n \eig{i}^{(k)} = z_n^{(k)} \eig{i}^{(k)}, \quad \LL \eig{i}^{(k)} = \lambda_i^{(k)} \eig{i}^{(k)}.
\end{equation}
We sort the eigenvalues in such a way that $\abs{\Re{\lambda_0}^{(k)}}<\abs{\Re{\lambda_1}^{(k)}} < \ldots < \abs{\Re{\lambda_n}^{(k)}}$.
The presence of a symmetry thus implies that the Liouvillian does not mix eigenoperators with different values of $k$, and therefore  the Liouvillian can be partitioned (block-diagonalized) into different symmetry sectors $\LL_{k}$, i.e., 
\begin{equation}
    \LL = \bigoplus_{k} \LL_{k}.
\end{equation}
For this reason, the eigenvalues ${\lambda_j}^{(k)}$ and eigenoperators $\eig{j}^{(k)}$ describe the whole physics within each of the Liouvillian symmetry sectors.

Weak symmetries fix the structure of the eigenoperators, which, on the number (Fock) basis, read
\begin{equation}\label{Eq:condition_symmetry}
    \eig{j}^{(k)}= \sum_{p,q} c_{p,q} \ket{p}\bra{q} \, , \quad \operatorname{mod}(p-q, n)=k,
\end{equation}
where $\operatorname{mod}(p-q, n)$ indicates the modulo operation.
In other words, $\eig{j}^{(k)}$ must be an operator containing only elements such that $(m-n)$ is either $k$, or $k\pm n$, or $k\pm 2n$, etc.
For example, for a $Z_2$ symmetry, this implies $(m-n)$ either even or odd, and therefore the eigenoperators of the Liouvillian must be characterized by a checkerboard-like structure.
We show a typical steady-state structure for a weak $Z_3$ and $Z_4$ symmetries in Figs.~\ref{fig:checkerboards}(a)~and~(b), respectively.
As also demonstrated in Ref.~\cite{AlbertPRA14}, in the case of a weak symmetry, $\sss$ is generally unique and thus \textit{must} belong to the $k=0$ symmetry sector of the Liouvillian. 
For this reason, for any finite number of photons in the system, the $ n$-photon-driven Kerr resonator with weak $Z_n$ symmetry will admit a unique steady state $\sss \propto \eig{0}^{(0)}$.

Furthermore, the discontinuous behavior of the steady state in \eqref{Eq:DPTDefinition} is signaled by the Liouvillian spectral properties. 
In the thermodynamic limit, a second eigenoperator, which is stationary under the action of the Liouvillian, emerges. 
Accordingly, an eigenvalue $\lambda_m^{(k)}$ becomes exactly zero, both in its real and imaginary parts, as a function of the parameter $\zeta$.
In finite-size systems, phase transitions cannot be observed, and $\lambda_m^{(k)}\neq 0$ if $m\neq 0$ and $k\neq 0$.
Nevertheless, the study of the Liouvillian spectral properties provides much useful information about the scaling and nature of the transition \cite{VicentiniPRA18}.

Within this formalism and notation, a first-order phase transition can be seen as a change in the $k=0$ symmetry sector, where the steady state $\sss\propto\eig{0}^{(0)}$ and the eigenoperator $\eig{1}^{(0)}$ display level touching (detail can be found in Ref.~\cite{MingantPRA18_Spectral}).
More specifically, $\lambda_1^{(0)}=0$ at the critical point, and the minimum in $\lambda_1^{(0)}$ reaches zero as the system scales towards the thermodynamic limit.

A spontaneous symmetry breaking of ${Z}_n$, instead, means the emergence of $n-1$ states, each one belonging to a different $k$-symmetry sector, that does not evolve anymore under the action of the Liouvillian.
In this case, the phase transition is associated with $\lambda_0^{(1)}, \dots \lambda_0^{(n-1)}$ becoming and remaining zero in a whole region where the symmetry is broken.
For instance, in the case of a $\mathcal{Z}_{2}$ breaking, $\lambda_0^{(1)}=0$ after the transition, while  for $\mathcal{Z}_{3}$ one has $\lambda_0^{(1)}=\lambda_0^{(2)}=0$.
The corresponding states $\eig{0}^{(k)}$, belonging to different symmetry sectors with respect to $\eig{0}^{(0)}$, allow constructing the symmetry-breaking steady states.
Indeed, by choosing the correct superposition of the form $\symmat{j}=\sum_{k}^{n-1} c_{i,k} \eig{0}^{(k)}$, one can obtain well-defined density matrices such that $\mathcal{Z}_n \symmat{j}\neq z_n^{(k)} \symmat{j}$ but $\LL \symmat{j} = 0$.

\subsubsection{DPTs and strong symmetries}

In the case of a strong symmetry, any eigenoperator is characterized by two quantum numbers $(k_{\rm L}, k_{\rm R})$, such that
$\mathcal{Z}_n^{\rm L, \, R} \eig{i}^{(k_{\rm L}, k_{\rm R})} = e^{\pm 2 i \pi k_{\rm L, \, R} /n}  \eig{i}^{(k_{\rm L}, k_{\rm R})}$, where, again, $k_{\rm L, \, R}=0, 1, \dots n$.
We deduce that 
\begin{equation}\label{Eq:condition_symmetry_srong}
\begin{split}
    \eig{i}^{(k_{\rm L}, k_{\rm R})}&= \sum_{p,q} c_{p,q} \ket{p}\bra{q} \, , \\
     \operatorname{mod}(p, n)=k_{\rm L}, & \quad \operatorname{mod}(q, n)=k_{\rm R}.
\end{split}    
\end{equation}
Such a structure is shown in Figs.~\ref{fig:checkerboards}(c)~and~(d) for two steady states of the $(0,0)$ symmetry sector for $Z_3$ and $Z_4$ symmetries.

Notice now that we can define two different types of eigenoperators: those which describe the evolution of populations, for which $k_{\rm L}=k_{\rm R}$, and the coherences, for which $k_{\rm L}\neq k_{\rm R}$.
Consequently, the symmetry sectors are $\LL_{k_{\rm L}, k_{\rm R}}$, i.e., 
\begin{equation}
    \LL = \bigoplus_{k_{\rm L}, k_{\rm R}} \LL_{k_{\rm L}, k_{\rm R}}.
\end{equation}
For each of the population sectors, there must exist a well-defined steady state $\sss^{(k)} \propto \eig{0}^{(k,k)}$(trace one, Hermitian, and positive semidefinite matrix which does not evolve under the action of the Liouvillian), while coherences are always traceless matrices.
Accordingly, the definition of the phase transition and the spontaneous symmetry breaking accounts for the presence of multiple disconnected eigenspaces.

A first-order DPT occurs in the population sectors, and it is associated with the presence of an eigenoperator $\eig{1}^{(k,k)}$ whose eigenvalue $\lambda_1^{(k,k)}$ becomes zero in the thermodynamic limit.
A spontaneous symmetry breaking, instead, implies that the eigenoperators $\eig{0}^{(k_{\rm L}, k_{\rm R})}$ acquire an eigenvalue $\lambda_1^{(k_{\rm L}, k_{\rm R})}=0$.
Spontaneous symmetry breaking thus implies that quantum superpositions between the states composing $\sss^{(k_{\rm L})}
\propto \eig{0}^{(k_{\rm L}, k_{\rm L})}$ and $\sss^{(k_{\rm R})}
\propto \eig{0}^{(k_{\rm R}, k_{\rm R})}$, i.e., two steady states of different symmetry sectors, can be maintained indefinitely.
Indeed, not only the populations do not evolve, but also the coherences remain stationary. 
In this regard, DPTs accompanied by spontaneous breaking of strong symmetries bear a closer resemblance with Hamiltonian transitions, and this is the reason for their use in quantum information \cite{PRLLieu20,gravina2022critical}.

\section{Semiclassical analysis of the 
$n$-photon driven resonator}
\label{sec:semiclassical}

The equation of motion for the expectation value of the observable $\hat{a}$ evolving under \eqref{lin} is
\begin{equation} \begin{split}
    \partial_t \expec{\hat{a}}_t &=  - i \sum_{m=1}^{m_{\rm max}} U_m \expec{\left(\hat{a}^\dagger \right) ^{m-1} \hat{a}^m}_t -i n G_n \expec{\left(\hat{a}^\dagger\right)^{n-1}}_t \\ & \quad  -\frac{\gamma}{2} \expec{\hat{a}} -\frac{n \eta_n}{2} \expec{\left(\hat{a}^\dagger \right) ^{n-1} \hat{a}^n}_t.
    \end{split}
\end{equation}
Due to the presence of non-quadratic terms, these equations of motion cannot be closed, leading to a hierarchy of coupled equations.

\subsection{The thermodynamic limit and finite-component phase transitions}
\label{subsec:thermodynamic}

We now introduce the dimensionless parameter $L$ such that  
\begin{equation}\label{Eq:Rescaling}
    G_n = \tilde{G}_n/\sqrt{L^{n-2}}, \, U_m = \tilde{U}_m/L^{m-1},  \, \eta_n = \tilde{\eta}_n/L^{n - 1}, 
\end{equation}
and we will consider the thermodynamic limit $L\to\infty$.
In such a limit 
$(G_n)^\alpha U_m$ and $(G_n)^\beta\eta_n$ are constants [for $\alpha$ and $\beta$ such that $(-n/2+1)\alpha-m+1=0$ and $(-n/2+1)\beta-n+1=0$], but the number of excitations diverges. Such a rescaling of the system parameters can be seen as the generalization of the scaling proposed in Ref.~\cite{CasteelsPRA17-2} for the $n=1$ case.
The semiclassical (coherent state) approximation amounts to assuming that the state of the resonator is coherent, i.e.,
\begin{equation}
\label{sc}
\rhot = \ket{\alpha(t)}\bra{\alpha(t)},
\end{equation}
where $\hat{a}\ket{\alpha(t)} = \alpha(t) \ket{\alpha(t)}$.
Accordingly, the equation of motion for the rescaled coherent field $\tilde{\alpha}(t)=\braket{\hat{a}} / \sqrt{L}$ leads to a generalized driven-dissipative Gross-Pitaevskii-like equation 
\begin{equation} \label{lin_sc}
\begin{split}
    \partial_t \tilde{\alpha} &=  \left[- i \sum_m \tilde{U}_m |\tilde\alpha|^{2(m-1)}    -\frac{n}{2} \tilde{\eta}_n|\tilde\alpha|^{2(n-1)} \right. \\ & \quad \left.  -\frac{\gamma}{2} \right] \tilde\alpha  -i n \tilde{G}_n \left(\tilde\alpha^* \right)^{n-1}.
    \end{split}
\end{equation}
Equation \eqref{lin_sc} is independent of $L$ and the photon number scales as $N=|\alpha|^2\propto L$ confirming that $L\to\infty$ corresponds to a well defined thermodynamic limit with an infinite
number of photons.
The parameter $L$ allows introducing the idea of finite-component phase transitions --- where the thermodynamic limit is replaced by a scaling of the system parameters \cite{CasteelsPRA16,CasteelsPRA17,CasteelsPRA17-2,BartoloPRA16,peng_unified2019,felicetti2020universal,minganti2021liouvillian,HwangPRA18}.
In general, we expect the semiclassical approximation (\ref{sc}) to be valid and predictive in the $L\to\infty$ limit, and far from the critical points where nonlinear processes inducing quantum fluctuations cannot be neglected.
This assumption, corroborated by the previously-cited extensive literature corpus, bears resemblance to the mean-field approximation in all-to-all connected two-level systems, where a similar approximation becomes valid in the limit of an infinitely large number of systems \cite{CarolloPRL21,HuybrechtsPRB20}.

\subsection{Analysis of the transition properties}

Given the invariance of \eqref{lin_sc} to the transformations in \eqref{Eq:Rescaling}, in the following analysis we will work with the {\it bare} quantities $\{U_m,\eta_n,G_n\}$.

Despite the simplification introduced by the semiclassical approximation, \eqref{lin_sc} cannot be yet analytically solved.
At the steady state, i.e.,  $\partial_t \alpha=0$,
\eqref{lin_sc} reads
\begin{equation}
\label{ss_sc}
      \left(     \sum_{m=1}^{m_{\rm max}} U_m N^{m-1}    - i \frac{\gamma + n \eta_n N^{n-1}  }{2} \right) \alpha =   n G_n \left(\alpha^* \right)^{n-1}.
\end{equation}
In general, \eqref{ss_sc} gives rise to multiple solutions for the photonic field $\alpha$.
The onset of new \textit{stable} solutions of \eqref{ss_sc} can be associated with the emergence of phase transitions.
For example, when $n=1$ the cubic equation for $\alpha$ (obtained by considering $m_{\rm max} =2$) gives rise to the well-known S-shaped curve for the photon number \cite{Drummond_JPA_80_bistability} signaling the presence of a first-order phase transition (between a low- and high-density state) accompanied by a hysteresis region with multiple stable solutions in the thermodynamic limit \cite{CasteelsPRA17-2,BartoloPRA16,CasteelsPRA16}. 

If $n\geq 2$  \eqref{ss_sc} always admits the solution $\alpha =\alpha_{\rm vac} =0$.
However, the other solutions $\alpha$ of \eqref{ss_sc} cannot be analytically found. 
Our strategy is thus to solve an {\it inverse problem}. 
Being interested in studying the emergence of criticalities as the driving strength is varied, by multiplying both sides by their complex conjugate, one finally obtains the equation
\begin{equation}\label{Eq:G_fun_N}
    G_n(N) = \sqrt{\frac{4 \left(\sum_m U_m N^{m-1} \right)^2 + {\left( \gamma + n \eta_n N^{n-1} \right)^2}} {4 n^2 N^{n-2}}},
\end{equation}
where we selected the positive branch of the square root since, up to a phase,  one can always choose $G_n\in\mathbb{R}^+$ \footnote{This amounts to a change in the initial condition by sending $\hat{a} \to \hat{a}e^{i \varphi_0}$}.

\subsubsection{Second-order phase transitions and behavior around $\alpha=0$}
\begin{figure}[t!]
    \centering
    \includegraphics[width=0.48 \textwidth]{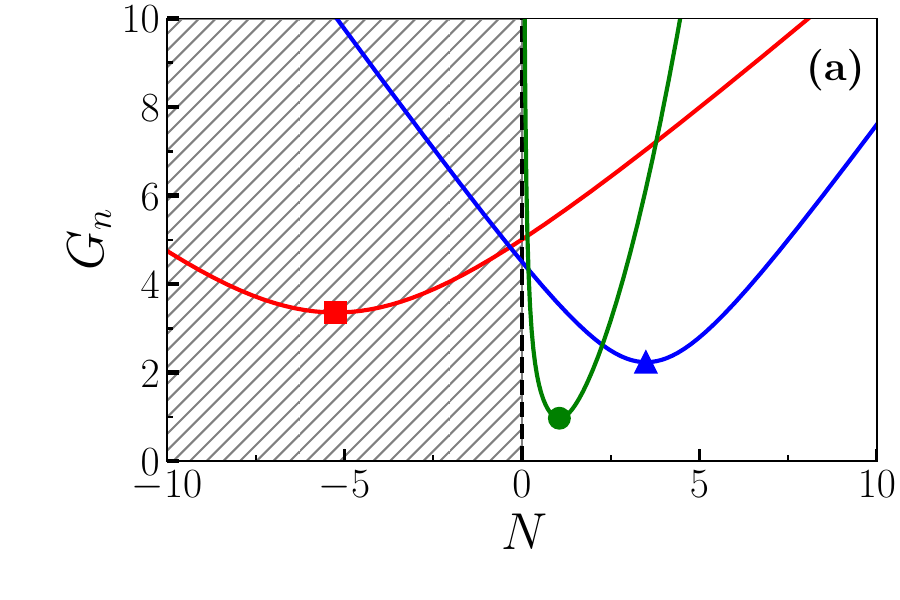}
        \includegraphics[width=0.48 \textwidth]{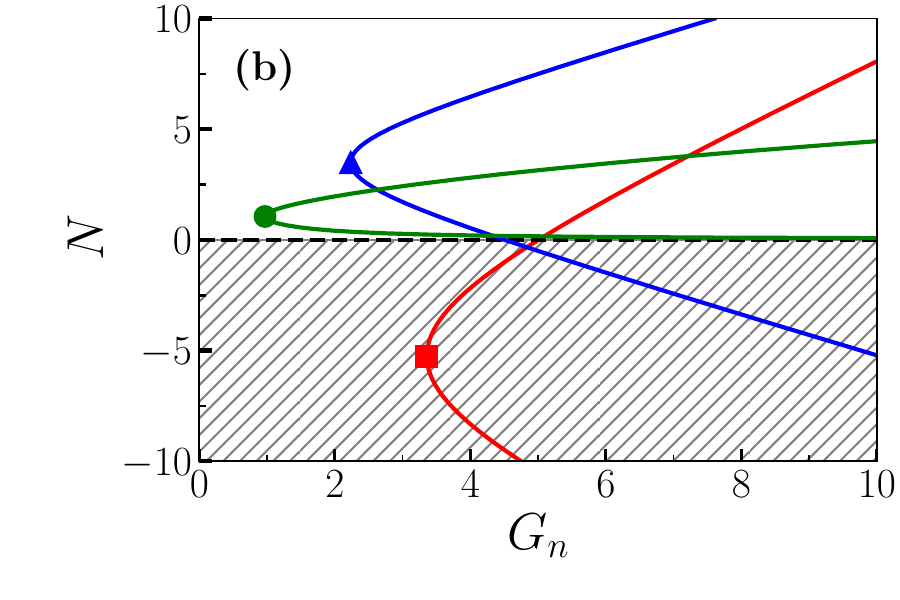}
    \caption{(a) Possible different behaviors of $G_n$ as a function of $N$ according to \eqref{Eq:G_fun_N}. The marker indicates the minima of the function $G_n$, while the hatching indicates the unphysical solutions $N<0$. (b) By simply inverting the plot, we can gain information on $N$ as a function of $G_n$.}
    \label{fig:SC_behavior}
\end{figure}
For the class of systems under consideration, a second-order phase transition occurs when the state changes from $N=\braket{\hat{a}^\dagger\hat{a}}=0$ to $N>0$ continuously as a function of the driving strength $G_n$ \cite{BartoloPRA16}. 
In other words, if a second-order DPT occurs, semiclassicaly the critical point must correspond to a solution of \eqref{Eq:G_fun_N} where
\begin{equation}
\label{def_critpoint}
G_n^{(c)}\equiv\lim_{N\to 0^+}G_n(N).
\end{equation}
We note that the limit $N\to0^+$ must be taken since \eqref{Eq:G_fun_N} is defined only for $N\neq0$. 
At this specific value of $G_n$ the system is thus allowed to pass from the semiclassical solution $\alpha_{\rm vac}=0$ to another stable solution with $\alpha \neq0$.

Notice that \eqref{Eq:G_fun_N} admits at most three possible behaviors around $N=0$ as sketched in Fig.~\ref{fig:SC_behavior}: 
\begin{itemize}
\item{The curve $G_n$ intersects the zero with a positive derivative (red line in Fig.~\ref{fig:SC_behavior}). In this case, the system can undergo a second-order DPT, passing continuously from the zero solution to a nonzero one.}
\item{The curve $G_n$ intersects the zero with a negative derivative (blue line in Fig.~\ref{fig:SC_behavior}).  This resembles the S-like shape of bistability in one-photon-driven systems. Since the photon number should monotonically increase by increasing the photon drive, this solution can never be stable and therefore the system can only undergo a first-order DPT.} 
\item{The curve $G_n$ never intersect the zero for a finite value of $N$ (green line in Fig.~\ref{fig:SC_behavior}). Also in this case the system can never experience a second-order DPT.} 
\end{itemize}

We conclude that a necessary (but not sufficient) condition to observe second-order DPTs, according to the semiclassical theory, is
\begin{subequations}
\begin{equation}\label{Eq:Condition_i}
    {\rm (i)} \quad 0 < G_n^{(c)}< \infty.
\end{equation}
\begin{equation}\label{Eq:Condition_ii}
   {\rm (ii)} \quad  \left.\frac{\partial G_n(N)}{\partial N}\right|_{G_n=G_n^{(c)}} \geq0.
\end{equation}
\end{subequations}
Notice that, in the case of a vertical-tangent point, higher-order derivatives need to be computed.

\subsubsection{Universal features and a semiclassical no-go theorem}

From the remarks in the previous section and using Eqs.~(\ref{Eq:Condition_i})~and~(\ref{Eq:Condition_ii}), we can already draw some important conclusions about the nature of DPTs in this class of systems. 
In particular, we formulate the following no-go theorem.

{\bf Semiclassical no-go theorem.}
{\it Consider a $n$-photon driven-dissipative resonator, with nonvanishing Kerr nonlinearity, governed by the Lindbladian (\ref{lin}). Then, according to the semiclassical equations of motion: (a) a second-order DPT never occurs for odd $n$; (b) If $\gamma \neq 0$ (weak symmetry),
$n=2$ is the only case with a second-order DPT; (c) If $\gamma = 0$, and $U_1 \neq 0$, again $n=2$ is the only cases where a second-order DPT can emerge; (d) A DPT for $n=4$  can be found only if $U_1 =\gamma = 0$.
(e) For $U_2 \neq 0$, no second-order DPTs can occur if $n>4$.} 
\begin{proof}
The semiclassical solutions of the stationary Gross-Pitaevskii equation (\ref{ss_sc}) must satisfy \eqref{Eq:G_fun_N}. The behaviour of this function around $N=0^+$ for $U_1\neq0$ or $\gamma\neq0$ is given by 
\begin{equation}
\label{expansion_1}
G_n(N)\simeq \frac{1}{2n} N^{\frac{2-n}{2}} \sqrt{4 U_1^2 + \gamma^2} \left[1+\mathcal{O}(N)\right].
\end{equation}
The case where $U_1,\gamma=0$, the expansion leads to 
\begin{equation}
\label{expansion_2}
G_n(N)\simeq\frac{|U_2|}{n} N^{\frac{4-n}{2}} \left[1+\mathcal{O}(N)\right].
\end{equation}

To prove \textit{(a)}, we consider odd-$n$, and
from \eqref{expansion_1} we get 
\begin{equation}
G_n^{(c)}=
\begin{cases}
0 \quad &\text{if $n=1$} \\
\infty \quad & \text{if $n=3,5,\dots$}
\end{cases}
\end{equation}
and therefore the condition (\ref{Eq:Condition_i}) for the occurrence of a second-order DPT is never satisfied.
If $U_1,\gamma=0$, instead, \eqref{expansion_2} gives
\begin{equation}
G_n^{(c)}=
\begin{cases}
0 \quad &\text{if $n=1,3$} \\
\infty \quad & \text{if $n=5,7,\dots$}
\end{cases}.
\end{equation}
We have therefore proven the statement \textit{(a)}.

Let us now consider the case of even $n$. 
From \eqref{expansion_1} we find that for $U_1\neq0$ or $\gamma\neq0$ a second order DPT is possible only for $n=2$, with a critical point given by
\begin{equation}
G_2^{(c)} = \frac{\sqrt{4U_1^2+\gamma^2}}{4}.
\end{equation}
Higher $n$ results in $G_n^{(c)} = \infty$.
Condition (\ref{Eq:Condition_ii}) reads
\begin{equation}
\label{2phcondition}
\left.\frac{\partial G_2(N)}{\partial N}\right|_{G_2=G_2^{(c)}} = \frac{U_1U_2+4\gamma\eta_2}{2\sqrt{4 U_1^2+\gamma^2}} 
\end{equation}
and thus it can be satisfied for an appropriate choice of the parameters.
These equations proves \textit{(b)} and \textit{(c)}.

\begin{figure}
    \centering
    \includegraphics[width = 0.49 \textwidth]{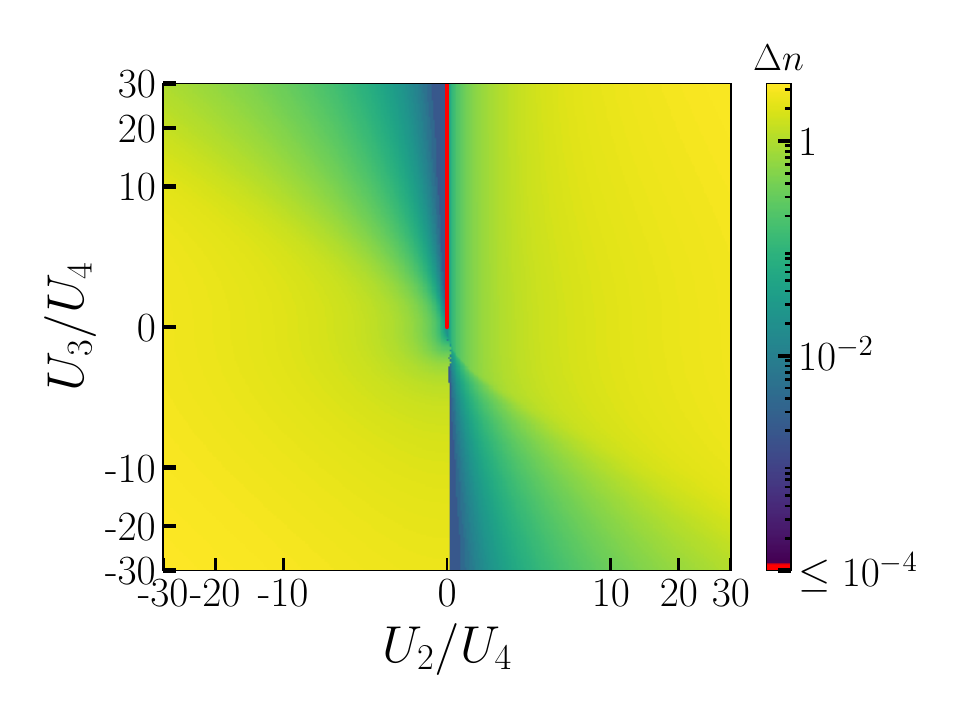}
    \caption{In a system with $n=6$ photon drive, analysis of the minimum possible jump in the photon number $\Delta n$ as a function of $U_2$ and $U_3$. This is defined in the main text as the stable semiclassical state with the smallest nonzero amplitude for all values of $G_6$. Parameters: $\gamma =0$, $U_1 = 0$, and $\eta_4/U_4 =0.1$.}
    \label{fig:six_photon_SC}
\end{figure}

Let us now assume $U_1=\gamma=0$. From \eqref{expansion_2} follows that, for $n=4$, a second-order DPT can can place also for
\begin{equation}
G_4^{(c)}= \frac{|U_2|}{4}.  
\end{equation}
Therefore, condition (i) in \eqref{Eq:Condition_i} is satisfied in the case $U_1=\gamma=0$. As for condition (ii) in \eqref{Eq:Condition_ii}, we have
\begin{equation}
\left.\frac{\partial G_4(N)}{\partial N}\right|_{G_4=G_4^{(c)}} =  \frac{U_3}{4}  \operatorname{Sign}\left(U_2\right),
\end{equation}
which can be satisfied choosing $U_2$ and $U_3$ with the same sign.
Finally, one can easily show that for $n>4$, \eqref{expansion_2} gives $G_n(N) = \infty$, demonstrating the impossibility of a DPT, thus proving \textit{(e)}.
\end{proof}

Before dealing with the analysis of the full quantum results, let us remark that $\gamma = 0$ is impossible to achieve in actual realizations. For many practical purposes, however, one can consider system ``sizes'' $L$ where, to a reasonable approximation, the role of $\gamma$ can be neglected, and thus the approximation $\gamma=0$ faithfully recovers the results of finite-time experiments.
Furthermore, the detuning terms can be easily manipulated, therefore making it possible to approximately fulfill the condition $\gamma=U_1 = 0$ necessary to witness the second-order DPT for the $n=4$ case.

We also notice that the mechanism enabling second-order DPTs for $n=4$ (i.e., the fact that $U_2$ and $U_3$ have the same sign) is the same behavior displayed by the two-photon Kerr resonator, where this role is played by the detuning $U_1$ and the two-photon interaction potential $U_2$ in \eqref{2phcondition}.
Finally, we stress that, although second-order DPTs could, in principle, also emerge for even $n>4$, these would require setting $U_2 = 0$, which, contrarily to detuning $U_1$, cannot be easily manipulated.

To demonstrate this fact, we compute $\Delta n$, defined as the minimal possible jump between the vacuum and a non-zero stable semiclassical solution, as a function of both $U_2$ and $U_3$.
To find that state, one spans all values of $G_n$ and finds the stable state with the minimal number of photons.
The results, reported in Fig.~\ref{fig:six_photon_SC} for $n=6$, show that for all nonzero $U_2$, the system always displays a finite jump associated with a first-order DPT, confirming that the system has no second-order transition.

\begin{figure}
    \centering
    \includegraphics[width=0.48 \textwidth]{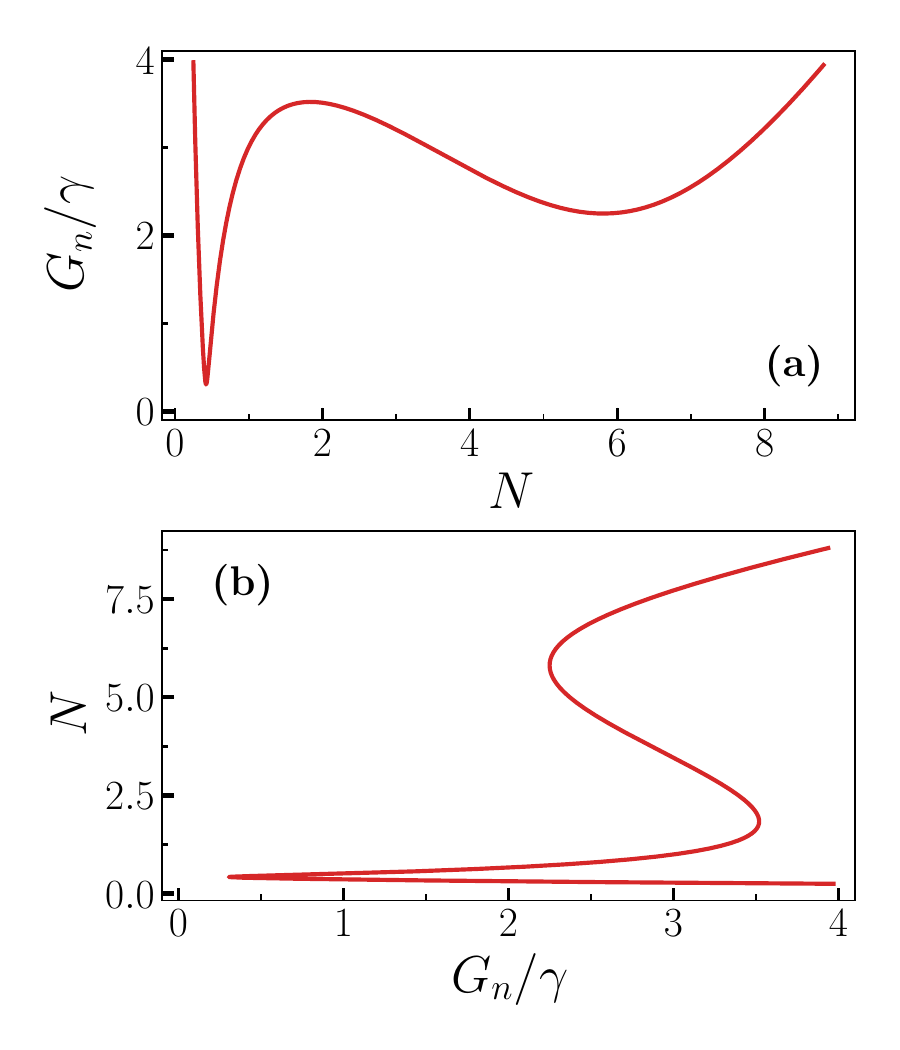}
    \caption{Multistability according to the semiclassical analysis in a four-photon driven resonator ($n=4$), where we fixed 
    $U_1=10 \gamma$, $U_2=-25 \gamma$, $U_3=3 \gamma$, $\eta_4=0.1 \gamma$.
    }
    \label{fig:multiple_first_order}
\end{figure}

\subsubsection{Stability of the vacuum across a first-order DPT and convergence radius of the semiclassical solution}
\label{subsubsec:vacuum_stability}
In this section, we show that, within the semiclassical picture, the solution $\alpha_{\rm vac} =0 $ is always asymptotically stable for $n>2$ if $\gamma \neq 0$ (i.e., in the presence of a first-order DPT).

Consider $\alpha=\alpha_{\rm vac} + \delta\alpha$, where $\delta\alpha\in\mathbb{C}$ is a small perturbation ($|\delta\alpha|\ll1$) around the vacuum solution.
Plugging the above parametrization into \eqref{lin_sc}, and expanding it at the first order in $\delta\alpha$, we get 
\begin{equation}\label{stabmat}
    \partial_t ({\delta\vec\alpha}) = \mathsf{M}\cdot {\delta\vec\alpha},
\end{equation}
where $\delta\vec\alpha = ({\rm Re}[\delta\alpha],  {\rm Im}[\delta\alpha])^\intercal$ and 
\begin{equation}\label{stabmat_def}
\mathsf{M}=
    \begin{pmatrix}
-\gamma/2 & -2\delta_{n,2}G_2+U_1 \\
-2\delta_{n,2}G_2-U_1 & -\gamma/2 
\end{pmatrix}
\end{equation}
is the so-called stability matrix. 
The solutions of \eqref{stabmat} are given by $\delta\vec\alpha(t)=\exp(-\lambda_\pm t)\delta\vec\alpha(0)$, where $\lambda_\pm=-\gamma/2\pm \sqrt{4\delta_{n,2}(G_2)^2-U_1^2}$ are the eigenvalues of $\mathsf{M}$. 
Thus, it is straightforward to conclude that for $n>2$ the vacuum solution $\alpha_{\rm vac}=0$ is always stable at a semiclassical level for finite single-photon losses since 
\begin{equation}
    {\rm Re}[\lambda_\pm]=-\frac\gamma2 < 0.
\end{equation}
For $n=2$ the vacuum gets unstable when 
\begin{equation}\label{Eq:condition_vacuum_unstable}
    {\rm Re}\left[\sqrt{4(G_2)^2-U_1^2}\right]>\frac\gamma2,
\end{equation}
which implies ${\rm Re}[\lambda_+]>0$.
Equation (\ref{Eq:condition_vacuum_unstable}) has significant consequences since it implies that, contrary to the $n=1,2$ case, the semiclassical dynamics never triggers a transition from the vacuum to high-density solutions if $\gamma \neq 0$. 
However, as we will see in  Secs.~\ref{sec:3ph_quantum_kerr} and \ref{sec:4ph_quantum_kerr}, quantum fluctuation in finite-size systems can make the vacuum solution unstable and allow for the onset of phase transitions. 

Finally, we note that in the case of strong symmetry $\gamma=0$, the vacuum is marginally stable, and higher-order perturbation theory is needed to assess the stability of the vacuum.

This stability analysis evaluates the stability of the vacuum solution to a weak perturbation (within the linear approximation)
but provides no information about the effect of non-infinitesimal perturbations of the vacuum. 
To this extent, one can perform an explicit numerical study of \eqref{lin_sc} considering a set of initial states $\alpha(t=0) = r \, e^{i \theta}$ (with $r\in\mathbb{R}^+$ and $\theta\in[0,2\pi]$), where we scan both $r$ and $\theta$.
One then studies the evolution of $\alpha(t)$, which can either converge back to the vacuum or reach one of the other stable solutions.
One then defines the convergence radius $r_{\rm max}$ as the maximal value of the radius $r$ for which the initial solution converges to the vacuum for all the possible values of the phase $\theta$ (see Sec.\ref{sec:semivsq} for an explicit example).

\begin{figure}[t!]
    \centering    \includegraphics[width=0.48\textwidth]{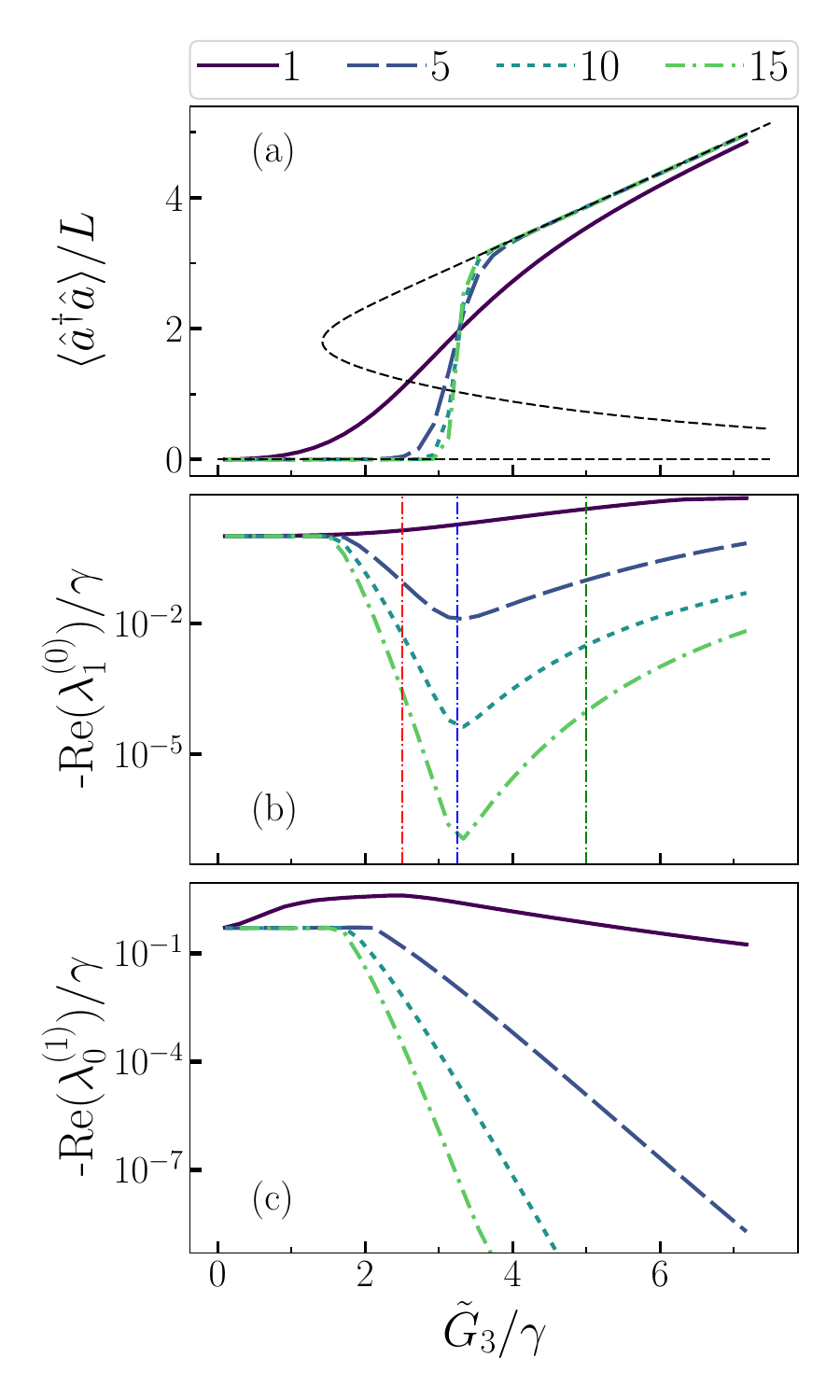}
    \caption{Onset of a first-order phase transition for increasing $L$ (see legend) with symmetry breaking in the three-photon Kerr resonator. Panel (a): mean number of photons in the steady state $\braket{\hat{a}^\dagger \hat{a}}$, renormalized by scaling parameter $L$. Panel (b): Real part of $\lambda_{1}^{(0)}$, i.e., the Liouvillian gap in the same symmetry sector as the steady state, inducing the first-order transition. The three vertical lines indicate the scaling values studied in Fig.~\ref{fig:scaling}. Panel (c): Real part of $\lambda_{0}^{(1)}$, i.e., the Liovillian eigenvalue signaling the spontaneous symmetry breaking.  Parameters: $U_1/\gamma=-20$, $U_2/\gamma=10$, $\eta_3/\gamma=1$.
    The cutoff $N_c$, also used in Figs.~\ref{fig:Eigendecomposition} and \ref{fig:scaling}, is chosen to ensure that, between the simulation with $N_c$ and $N_c+10$, all quantities differ by less than $1\%$.
    The cutoffs are: $N_c = 30$
    for $L=1$; $N_c = 95$ for $L=5$; $N_c = 140$ for for $L=10$; $N_c = 190$ for $L=15$. 
    }
    \label{fig:num_phot_and_gap_first}
\end{figure}

\subsubsection{Multistability of solutions with different number of photons}

The solution around $N=0$ predicts either a first- or a second-order phase transition describing the passage of the system from the vacuum to a nonzero population phase.
This analysis does not predict the behavior far from $N=0$, and nothing prevents several Hamiltonian terms from competing with each other, thus resulting in multiple stable solutions.
In particular, the presence of this multistability would imply an overlap of ``S-like'' curves of the semiclassical solution, so that for the same drive intensity, there are multiple solutions with different photon numbers.

To understand which mechanism can enable multistability, let us consider again \eqref{Eq:G_fun_N}.
In the semiclassical formalism, multistability implies the presence of multiple solutions at the semiclassical level  with different photon numbers [c.f. Fig.~\ref{fig:multiple_first_order}~(b)]. 
This translates in the presence of  multiple local minima (or maxima) of the function $G_n(N)$, as shown in Fig.~\ref{fig:multiple_first_order}~(a).
Therefore, one can study the function
\begin{equation}
\frac{\partial G_n(N)}{\partial N} =0.
\end{equation}
The number of maxima and minima signals the presence of multiple semiclassical solutions.
And following Descartes' rule of signs -- i.e.,  the maximal number of positive roots of a polynomial is the number of sign changes between consecutive coefficients -- we deduce that a necessary condition to have multiple solutions is the presence of alternating signs between the various $U_n$.
Physically speaking, the underlying mechanism is quite straightforward: different $U_n$ terms can compete with each other in determining the energy of one photon in the system, while drive and dissipation favor the solution with more or fewer photons. 
Since the relevance of each interaction term can change in different occupation regimes, several solutions can emerge.
The stability of the semiclassical solutions in the presence of quantum fluctuations needs to be numerically assessed.

\section{Three-photon Kerr resonator}
\label{sec:3ph_quantum_kerr}

Having discussed the general properties of DPTs, we turn now to specific examples to demonstrate the validity of the semiclassical analysis and show the quantum properties around criticality.
Throughout the next two sections, we will diagonalize the Liouvillian superoperator.
We take full advantage of the system's symmetry, as detailed in Appendix~\ref{App:method}, to reduce the computational complexity and enhance the precision of the results.
For the most numerically demanding simulations, we resort to the recently-developed Arnoldi-Lindblad method \cite{minganti2022arnoldi}, in conjunction with the algorithm detailed in Appendix~\ref{App:method}.

Here, we consider the three-photon-driven Kerr resonator governed by the master equation
\begin{equation}
\label{Eq:3ph}
\partial_t \hat{\rho}(t) = -i \left[\hat{H}_{3}, \hat{\rho}(t) \right] + \gamma \DD[\hat{a}] + \eta_3 \DD[\hat{a}^3]
\end{equation}
with
\begin{equation}
\hat{H}_{3} = U_1 \hat{a}^\dagger \hat{a} + \frac{U_2}{2} \left( \hat{a}^\dagger \right)^2 \hat{a}^2 + G_3 \left[\hat{a}^3  + \left(\hat{a}^\dagger\right)^3 \right].
\end{equation}
We focus on the $\gamma \neq 0$ case, where the system displays a $Z_3$ weak symmetry. 
According to the semiclassical analysis, we expect a first-order dissipative phase transition accompanied by the spontaneous breaking of the weak $Z_3$ symmetry.

\subsection{Semiclassical vs quantum solution}

First, we analyze the photon number as a function of the driving strength $G_3$.
In Fig.~\ref{fig:num_phot_and_gap_first}, we show the results of the full quantum analysis (colored lines) and compare them to the prediction of the semiclassical analysis (dashed black line). 
For a weak drive $G_3$, the system is in the vacuum, and $\sss \simeq \ket{0}\bra{0}$.
Increasing the drive intensity, the system's photon number deviates from the vacuum and approaches the high-photon number branch predicted by the semiclassical theory.
In this symmetry-broken phase, the stationary state is well-approximated by a statistical mixture of three coherent states, i.e., 
\begin{eqnarray}\label{Eq:approx_ss_in_3_phot}
\sss \simeq  \frac{\ket{\alpha_1}\bra{\alpha_1}+\ket{\alpha_2}\bra{\alpha_2} + \ket{\alpha_3}\bra{\alpha_3}}{3} ,
\end{eqnarray} 
where $\ket{\alpha_{1,2,3}}$ are coherent
states with the same number of photons  and a relative phase difference of $\pm 2\pi/3$, i.e.,
\begin{equation}
\alpha_{j+1} = \alpha_{j} \ e^{i \frac{2\pi}{3}}.
\end{equation}
The change in the steady state population becomes more and more abrupt as we increase the parameters $L$, demonstrating that, indeed, the phase transition is of the first order.

\begin{figure}
    \centering
    \includegraphics[width = 0.49 \textwidth]{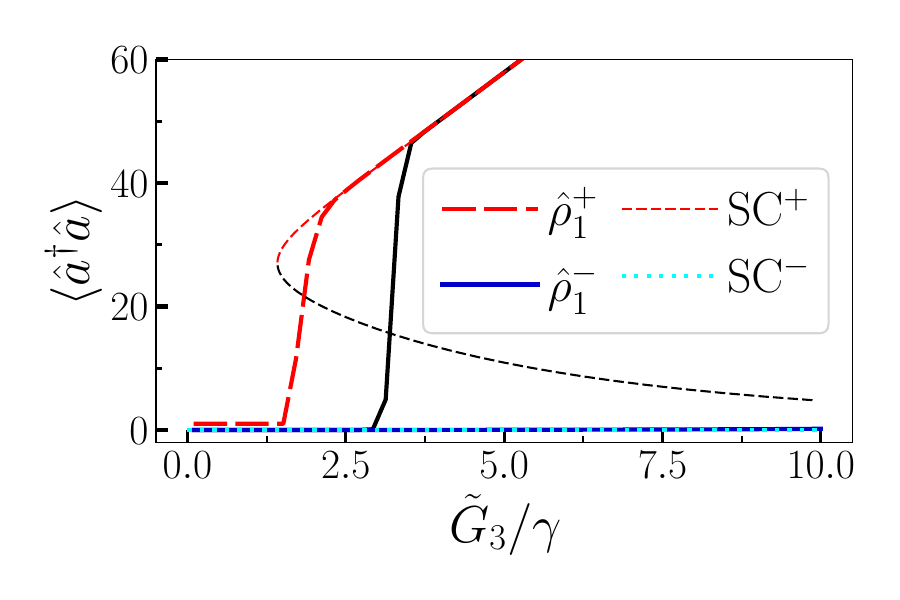}
    \caption{Eigendecomposition of $\eig{1}^{(0)}$ for $L =15$ and the parameters in Fig.~\ref{fig:num_phot_and_gap_first}. 
    The solid black line is the result of full quantum simulations. The semiclassical predictions are the dotted light blue curve (few photon numbers, stable), the dashed red curve (high-photon number, stable), and the  dashed black curve (unstable solution). The results of the eigendecomposition are plotted with solid blue and dashed red lines.}
    \label{fig:Eigendecomposition}
\end{figure}

\subsection{Analysis of the first-order transition}
\label{sec:semivsq}

To confirm the presence of a first-order phase transition, we plot in Fig.~\ref{fig:num_phot_and_gap_first}(b) the Liouvillian eigenvalue $\lambda_1^{(0)}$ associated with the slowest relaxation rate in the steady-state symmetry sector. This eigenvalue signals hysteresis and critical slowing down,
and the fact that it tends to zero in the thermodynamic limit proves the presence of a first-order DPT \cite{MingantPRA18_Spectral}.

We then investigate the properties of the eigenoperator $\eig{1}^{(0)}$ associated with such a state. According to Ref.~\cite{MingantPRA18_Spectral}, in the critical region one can use the eigendecomposition of $\eig{1}^{(0)}$ to recast
\begin{equation}
    \eig{1}^{(0)} \simeq \eig{1}^{+} - \eig{1}^{-},
\end{equation}
where $\eig{1}^{\pm}$ represent the density matrices of the metastable states.
As such, we expect that, in the thermodynamic limit, $\eig{1}^{\pm}$ recover the two stable solutions of the semiclassical theory.
We show the eigendecomposition in Fig.~\ref{fig:Eigendecomposition}. 
We indeed find that the semiclassical approximation qualitatively recovers the results of the eigendecomposition.

\begin{figure}[h]
    \centering
    \includegraphics[width = 0.49 \textwidth]{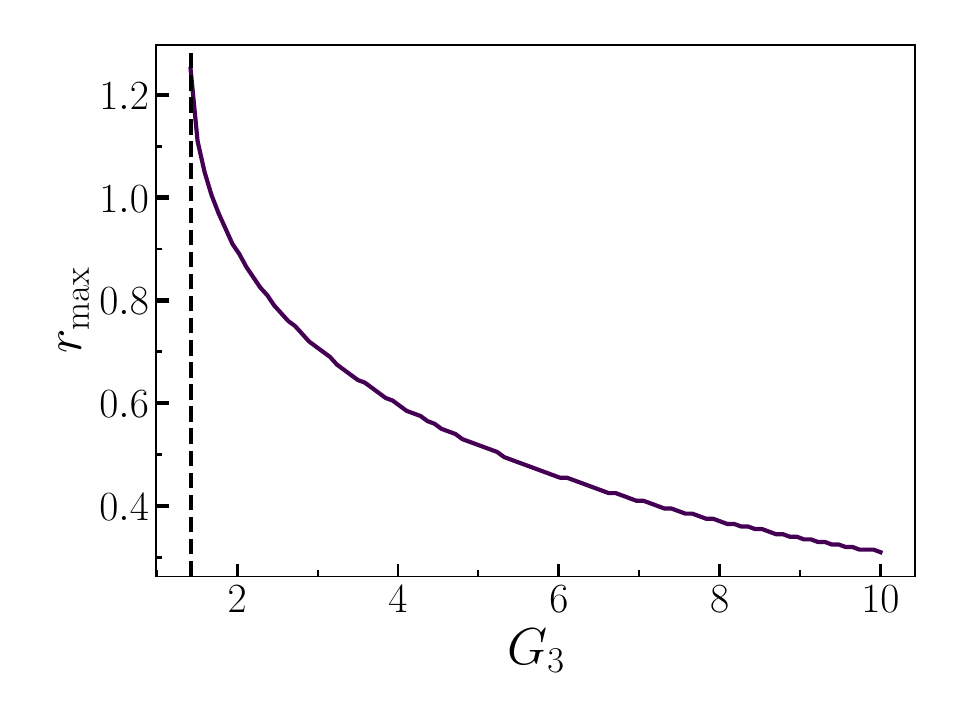}
    \caption{As a function of the drive amplitude, the convergence radius $r_{\rm max}$, numerically computed according to the procedure
    in Sec.~\ref{subsubsec:vacuum_stability}.
    The dashed vertical line indicates where the high-photon solution becomes stable. Below this value, the convergence radius becomes infinite.
    Parameters as in Fig.~\ref{fig:num_phot_and_gap_first}.
    }
    \label{fig:radius_3_photon}
\end{figure}

As discussed in Sec.~\ref{subsubsec:vacuum_stability}, the semiclassical analysis predicts the presence of a stable vacuum in the whole symmetry-broken region.
This is also shown in Fig.~\ref{fig:radius_3_photon}, where we plot the convergence radius $r_{\rm max}$ of the semiclassical approximation.
$r_{\rm max}$ remains finite for all values of the drive, despite it decreasing at larger drives.
Nonetheless, this decreases as a function of the pump amplitude $G_3$.
Within this picture, it is the presence of rare and collective quantum fluctuations that trigger the jump between the otherwise semiclassically stable solutions (faithfully representing the quantum state at both sides of the first-order transition). 
This phenomenology is typical of first-order DPTs (see, e.g., Refs.\cite{RodriguezPRL17,VicentiniPRA18}).

\begin{figure}
    \centering
    \includegraphics[width = 0.49 \textwidth]{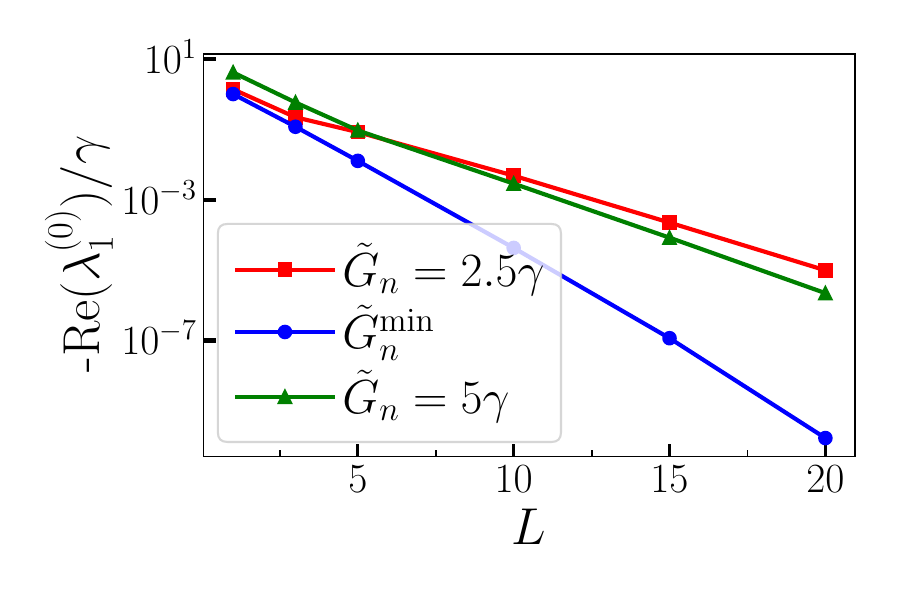}
    \caption{Scaling towards the thermodynamic limit for the three vertical lines in Fig.~\ref{fig:num_phot_and_gap_first}(b), demonstrating the presence of a first-order DPT and the emergent stability of the vacuum.}
    \label{fig:scaling}
\end{figure}

This analysis is confirmed both by the eigendecomposition in Fig.~\ref{fig:Eigendecomposition} (the vacuum remains long-lived even far from the transition point) and from the spectral analysis in Fig.~\ref{fig:num_phot_and_gap_first}(b). Considering larger values of $L$ results in slower timescales.
We confirm the scaling towards the thermodynamic limit of the Liouvillian gap $\lambda_1^{(0)}$ in Fig.~\ref{fig:scaling}. 
We consider a point before the transition (red line), at the minimum of the gap (blue line), and after the transition (green line). 
The same three lines correspond to the vertical lines in Fig.~\ref{fig:num_phot_and_gap_first}(b).
In all the cases, after an initial transient, we see an exponential closure of the gap as a function of $L$.
The green curve confirms the vacuum metastability predicted by  the semiclassical theory.

\subsection{Spontaneous symmetry breaking}

The spontaneous symmetry breaking implies that, for strong enough pumping, each of the state $\ket{\alpha_i}\bra{\alpha_i}$ in \eqref{Eq:approx_ss_in_3_phot} becomes a steady state of the system, since they are not eigenstates of $\mathcal{Z}_3$ \cite{MingantPRA18_Spectral}.
We confirm this picture in Fig.~\ref{fig:num_phot_and_gap_first}(c), where we show that also $\lambda_0^{(1)}$ becomes zero.
As expected, we obtain an identical result for $\lambda_0^{(2)}$ (not shown).
This implies that the system's coherent state has become metastable.

\begin{figure*}[t!]
    \centering
    \includegraphics[width = 0.98 \textwidth]{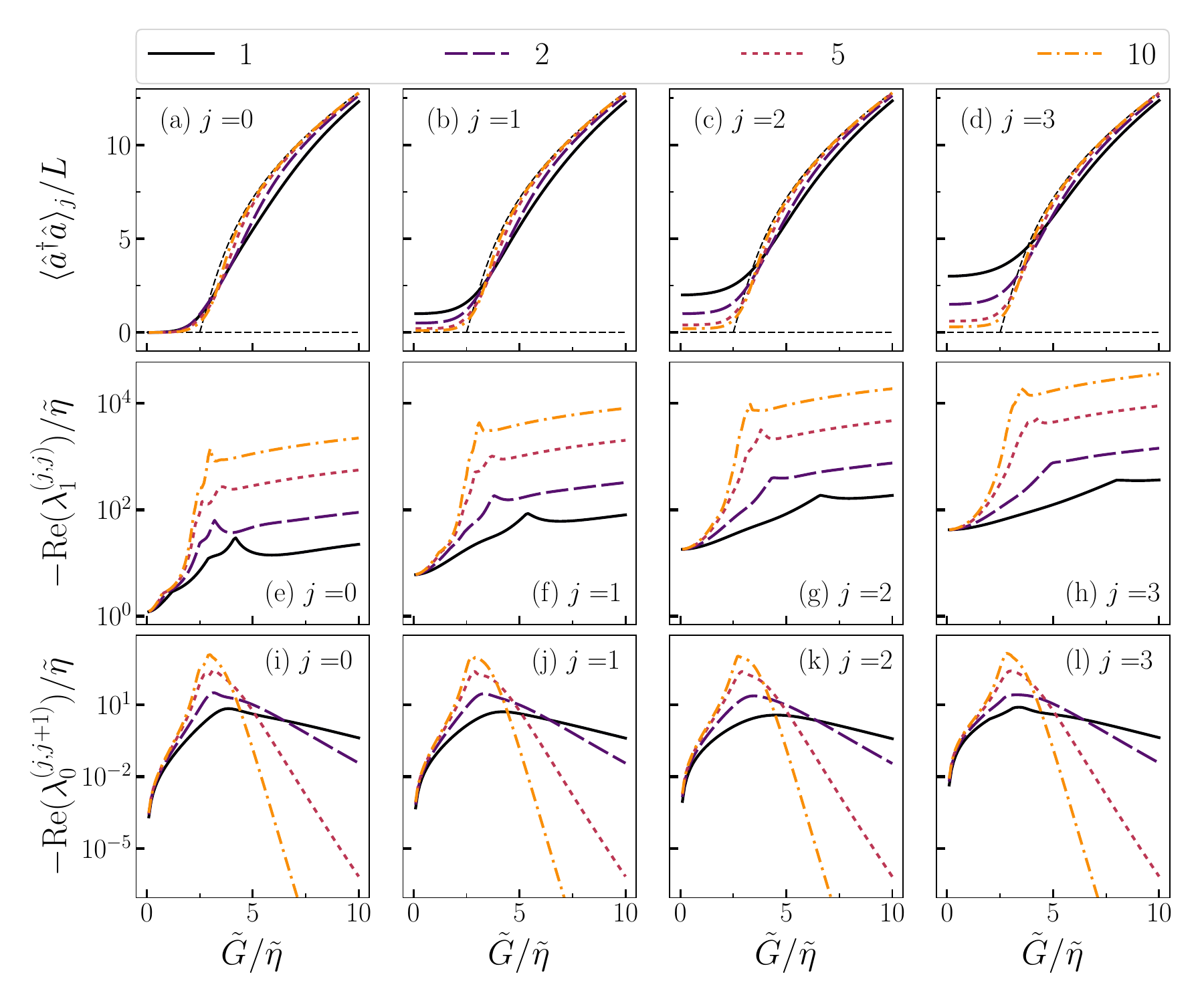}
    \caption{Study of the strongly-symmetric four-photon Kerr resonator, and the onset of a second-order dissipative phase transition.
    For different values of the thermodynamic rescaling parameter $L$:
    (a-d) photon number in the $(j,j)$ symmetry sector;
    (e-h) Liouvillian gap in the $(j,j)$ symmetry sector;
    (e-h) smallest Liouvillian eigenvalue in the $(j,j+1)$ symmetry sector.
    Parameters: $\gamma=U_1=0$, $U_2=10 \eta_4$, $U_3=\eta_4$. The cutoff $N_c$ is chosen to ensure that, between the simulation with $N_c$ and $N_c+10$, all quantities differ by less than $1\%$.
    The cutoffs are: $N_c = 45$
    for $L=1$; $N_c = 75$ for $L=2$; $N_c = 150$ for for $L=5$; $N_c = 290$ for $L=10$. 
    }
    \label{fig:four_photon}
\end{figure*}

\section{Four-photon Kerr resonator}
\label{sec:4ph_quantum_kerr}

Here, we consider the four-photon driven-dissipative Kerr resonator, reading
\begin{equation}
\label{Eq:4ph}
\partial_t \hat{\rho}(t)= -i \left[\hat{H}_{4},\hat{\rho}(t)  \right] + \gamma \DD[\hat{a}] + \eta_4 \DD[\hat{a}^4]
\end{equation}
with 
\begin{equation}
\begin{split}
\hat{H}_{4} = U_1 \hat{a}^\dagger \hat{a} & + \frac{U_2}{2} \left( \hat{a}^\dagger \right)^2 \hat{a}^2 + \frac{U_3}{3} \left( \hat{a}^\dagger \right)^3 \hat{a}^3 \\
&   + G_4 \left[\hat{a}^4  + \left(\hat{a}^\dagger\right)^4 \right].
\end{split}
\end{equation}

\subsection{Strong symmetry and second-order phase transition}

We start by considering the strong symmetric case $\gamma = 0$ with $U_1 =0$.
For this set of parameters, the semiclassical analysis predicts a second-order phase transition associated with the spontaneous breaking of a $Z_4$ strong symmetry. We analyze it in Fig.~\ref{fig:four_photon}.
We recall that since the system has a strong $Z_4$ symmetry, the number of Liouvillian sectors is $4 \times 4$, being characterized by the two quantum numbers $k_{\rm L}$ and $k_{\rm R}$. 
The $4$ sectors with $k_{\rm L}=k_{\rm R}$ describe the evolution of the populations, while the remaining $12$ with  $k_{\rm L}\neq k_{\rm R}$ describe the evolution of the coherences.

First, we consider the re-scaled photon number of the steady state for each of the  symmetry sectors $(j,j)$ with $j \in [0, 3]$ and increase the thermodynamic parameter $L$.
Calling $\sss^{(j)} \propto \eig{0}^{(j,j,)}$ the steady state in each symmetry sector,
 $\langle \hat{a}^\dagger \hat{a} \rangle_j  = \Tr{\hat{a}^\dagger \hat{a} \sss^{j}}$ are
 plotted in Figs.~\ref{fig:four_photon}(a-d).
For low drive amplitudes, the system is in the $Z_n$ symmetric vacuum.
Indeed, the states need to respect the strong symmetry condition in \eqref{Eq:condition_symmetry_srong}, and thus 
\begin{equation}
    \sss^{(j)} = \ket{{\rm vac}_j}\bra{{\rm vac}_j} = \ket{j}\bra{j}, 
\end{equation}
where $j$ labels the symmetry sector and $\ket{j}$ is the Fock state with $j$ photons.
For large drive, instead, the system transition towards 
\begin{equation}
\label{Eq:4_cat}
\sss^{(j)} \simeq \ket{\mathcal{K}_j}\bra{\mathcal{K}_{j}}
\end{equation}
where the Schr\"odinger cats $\ket{\mathcal{K}_{i}}$ are
\begin{equation}
    \ket{\mathcal{K}_j} = \frac{1}{\mathcal{N}}\sum_{n=0}^3 e^{i\pi jn/2} \ket{\alpha_n},
\end{equation}
where $\ket{\alpha_n}$ are coherent states such that $\alpha_n=e^{i\pi n/2} \alpha$
and $\mathcal{N}$ is a normalization factor.
Increasing the value of $L$ towards the thermodynamic limit, we observe that the passage between the $Z_n$ vacua and the cat states becomes sharper and sharper, but remains continuous. 
This analysis corroborates the semiclassical one, and by appropriately taking into account the system's symmetry, we observe a second-order DPT.
To further demonstrate that, indeed, the transition is of the second and not of the first order, we plot $\lambda_1^{(j,j)}$ in Figs.~\ref{fig:four_photon}(e-h), i.e., the Liouvillian gap of the $(j,j)$ symmetry sector.
We observe no closure of the Liouvillian gap, indicating that no critical slowing down or hysteresis occurs for the Liouvillian populations.

Finally, we plot the smallest Liouvillian eigenvalue $\lambda_0^{(j,j+1)}$ for the sectors $(j,j+1)$ (where $j+1 = 0$ if $j=3$) in Figs.~\ref{fig:four_photon}(i-l).
These represent the decay rate of coherences between the sector $j$ and $j+1$, and their closure indicates the possibility of retaining everlasting coherences.
In this case, we observe that, after the critical point, these eigenvalues progressively become smaller, indicating that the system undergoes a second-order phase transition.
We obtain similar results for the other $(j,k)$ sectors with $j \neq k$ (not shown).
This is associated with a spontaneous breaking of the strong $Z_4$ symmetry, because it results in 
\begin{equation}
\begin{split}
 & \hat{\rho} \propto \left(\ket{\mathcal{K}_j} +  \ket{\mathcal{K}_k}  \right) \left(\bra{\mathcal{K}_j} +  \bra{\mathcal{K}_k}  \right), \\
 & \LL \hat{\rho} = 0 \quad{\rm but} \quad  \mathcal{Z}_4^{\rm L, R} \hat{\rho} \neq z_4^{\rm L, R}  \hat{\rho}.
 \end{split}
\end{equation}

\subsection{Weak symmetry and multistability}

\begin{figure*}[t]
    \centering
    \includegraphics[width = 0.98 \textwidth]{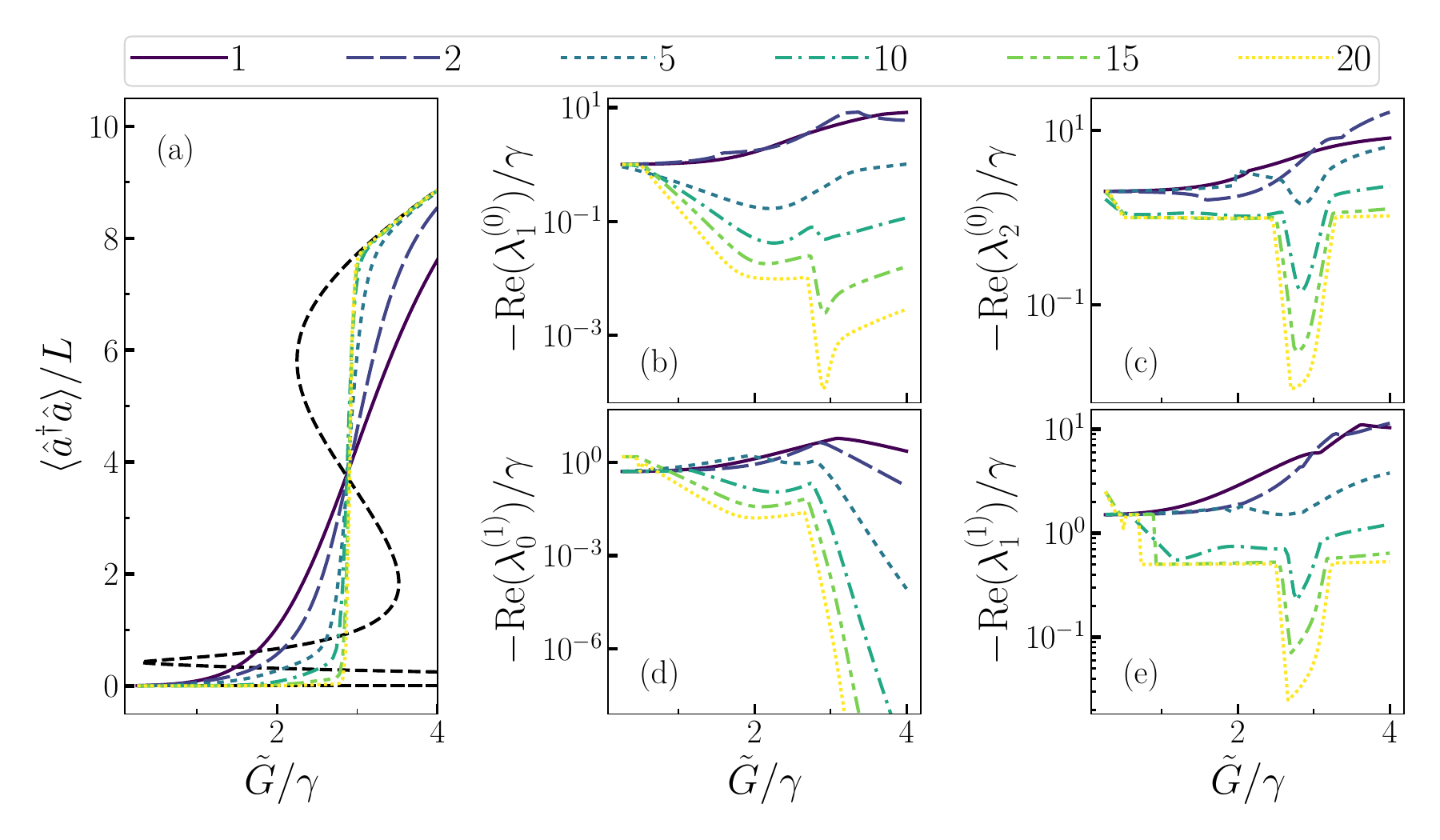}
    \caption{Analysis of the classically multistable system.
    (a) Photon number as a function of the drive for different values of the thermodynamic scaling parameter $L$. The black dashed line represents the semiclassical solution.
    (b) Liouvillian gap and (c) second Liouvillian eigenvalue in the $k=0$ sector, demonstrating the presence of two slow timescales.
    (d) Smallest (e) and second smallest Liouvillian eigenvalues in the $k=1$ sector, demonstrating the presence of SSB and of a slow timescale.
    Parameters: $U_1=10 \gamma$, $U_2/\gamma=-25 \gamma$, $U_3=3 \gamma$, $\eta_4=0.1 \gamma$.
    The cutoff $N_c$, also used in Fig.~\ref{fig:eigendecomposition_tristability}, is chosen to ensure that, between the simulation with $N_c$ and $N_c+10$, all quantities differ by less than $1\%$.
    The cutoffs are: $N_c = 40$
    for $L=1$; $N_c = 60$ for $L=2$; $N_c = 100$ for $L=5$; $N_c = 190$ for $L=10$; $N_c = 280$ for $L=15$; $N_c = 320$ for $L=20$. 
    }
    \label{fig:tristability}
\end{figure*}

We now consider a weakly symmetric case in the presence of detuning $U_1$ and with competing terms giving rise to multistability according to the semiclassical solution.
First, in Fig.~\ref{fig:tristability}(a), we compare the results of the semiclassical analysis with those of the full quantum simulation.
We find that, although the semiclassical solution has three stable solutions, the full quantum simulation is characterized by  a single first-order DPT, from the vacuum to the highest-populated manifold.
Indeed, if we analyze the Liouvillian gap $\lambda_1^{(0)}$ in Fig.~\ref{fig:tristability}(b) we see the closure of the Liouvillian gap associated with a first-order DPT.
If, however, we also consider the second eigenvalue $\lambda_2^{(0)}$  as in Fig.~\ref{fig:tristability}(c), we see that a second slow timescale emerges. That is, despite the presence of a single phase transition, the dynamics of the population of the system are characterized by \textit{two slow timescales}.

We corroborate this phenomenon by analyzing the symmetry sectors responsible for  spontaneous symmetry breaking.
In Fig.~\ref{fig:tristability}(c), we plot $\lambda_0^{(1)}$ showing that, indeed, this phenomenon is accompanied by the breaking of the weak $Z_4$ symmetry.
Noticeably, the spontaneous symmetry breaking takes place before the occurrence of the first-order transition.
Furthermore, we also observe a second slow timescale for this symmetry sector, i.e., $\lambda_1^{(1)}$ in Fig.~\ref{fig:tristability}(d).
These slow timescales represent the fact that there exist multiple symmetry-broken states, and there is a slow rate at which the system switches between them.
We observe similar results for the other symmetry sectors (not shown).

The picture we derive is one in which, although there are only two real steady states of the dynamics, either the vacuum or the one at large photon number, there exists a third metastable state to which the system can be initialized. 
Such a state is characterized by a broken symmetry, but it cannot be reached by quantum fluctuation alone.

\begin{figure}[t!]
    \centering
    \includegraphics[width = 0.49 \textwidth]{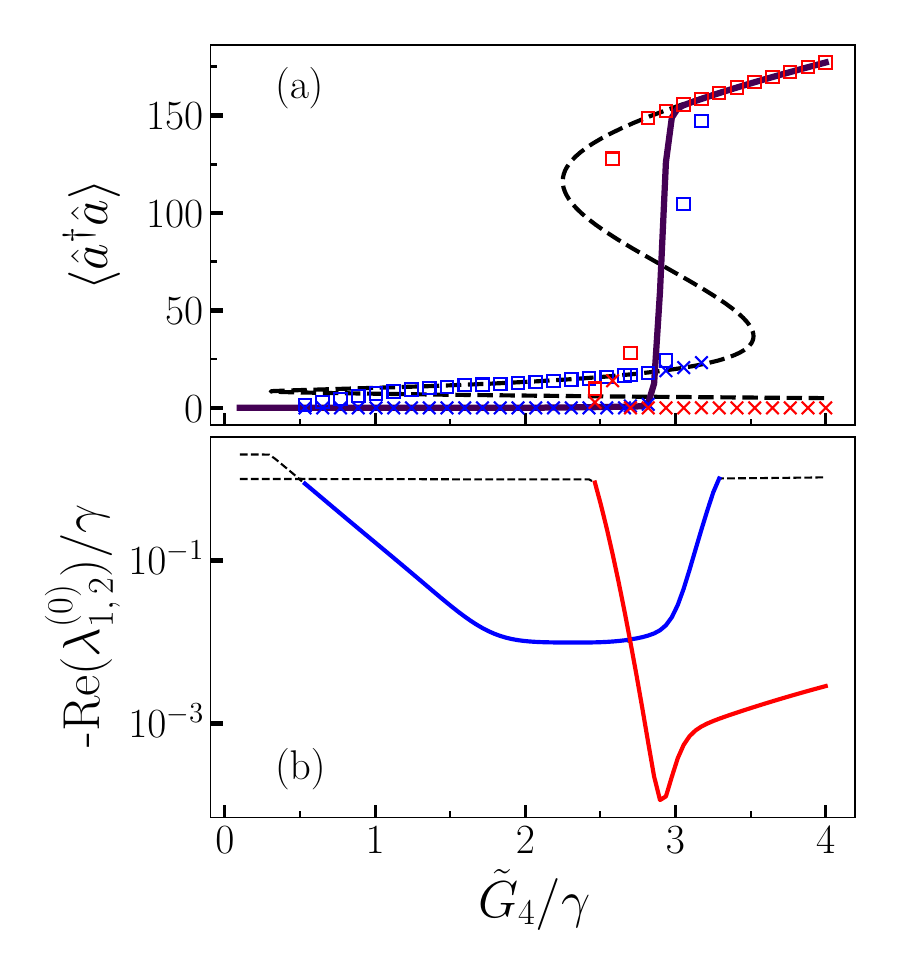}
    \caption{Eigendecomposition and comparison with the semiclassical solution.
    (a) Photon number of the full quantum solution (black solid line) compared to the semiclassical solution (black dashed line) and the results of the eigendecomposition [red and blue markers correspond to the red and blue curves in panel (b)]. 
    (b) The two smallest Liouvillian eigenvalues, whose behavior across the transition has been reconstructed using the continuity of the associated eigenoperators.
    Parameters as in Fig.~\ref{fig:tristability} for $L=20$.
    }
    \label{fig:eigendecomposition_tristability}
\end{figure}

To further demonstrate this picture, in Fig.~\ref{fig:eigendecomposition_tristability}(a) we use the eigendecomposition to express the eigenoperators associated with the slowest eigenvalues as
\begin{eqnarray}
    \eig{1}^{(0)}= \eig{1}^{(0), +} - \eig{1}^{(0), -}, \quad \eig{2}^{(0)}= \eig{2}^{(0), +} - \eig{2}^{(0), -}.
\end{eqnarray}
As one can see, these metastables density matrices recover the results of the semiclassical analysis, and the region in which there is a closure of these Liouvillian eigenvalues roughly corresponds to the region of multistability according to the semiclassical analysis.

Overall, the system displays 9 metastable coherent-like states, approximated by $\ket{\alpha_{\rm vac}}$, $\ket{\alpha_{\rm low} e^{i \phi_j}}$, and $\ket{\alpha_{\rm high} e^{i \phi_j}}$, with $|\alpha_{\rm vac}|<|\alpha_{\rm low}|<|\alpha_{\rm high}|$ and $\phi_j \in j \pi / 4 $.

\section{Conclusions and outlook}
\label{sec:con}

In this work, we explored the critical properties of $n$-photon driven-dissipative nonlinear quantum resonators.
We found that the symmetries of the model, fixed by driving and dissipation, determine the nature of the phase transitions in the steady state.
We characterize such criticalities providing general results for this class of models. 

We attack the problem using a semiclassical approach valid in a well-defined thermodynamic limit with an infinite number of excitations. In such a limit the state of the system approaches a coherent state and quantum fluctuations are suppressed leading to a generalized version of the driven-dissipative Gross-Pitaevskii equation. 
Studying its stationary properties we formulate and prove a no-go theorem stating that no second-order phase transitions are possible when $n$ is odd, while, for even $n$, second-order transitions can take place only for $n=2$ and $n=4$.  

We then perform a full quantum analysis of the three- and four-photon-driven Kerr resonators. We find that quantum fluctuations trigger the transition between semiclassical solutions in the thermodynamic limit validating the results obtained in the semiclassical limit. 
While the semiclassical approximation has been proved to be reliable for $n=1,2$, for higher $n$ there are no strong arguments supporting its validity.
Indeed the systematic inclusion of small quantum fluctuation on top of the mean-field semiclassical solution can be obtained via truncated Wigner methods \cite{Risken1967,RiskenPRA87} and Gaussian expansions \cite{VerstraelenPRR20}.
This is not the case for $n>2$ because the drive and $n$-photon dissipation  can, in principle, introduce non-Gaussian correlation above the coherent-state solution \cite{MingantiPRA23}.
The emergence of these dissipative phase transitions is understood and characterized within the spectral theory of Liouvillian highlighting the role of weak and strong symmetries. 

These results could also be relevant in the field of quantum technologies and quantum information encoding.
Symmetry breaking in second-order DPTs has been demonstrated to be a resource to improve the sensitivity of quantum measurement protocols \cite{di2021critical,HeugelPRL19}.
Our work proves that such kind of enhancement can only be attained for $n=2$ or $n=4$. 
Furthermore, our results may pose constraints for the exploitation of nonlinear-driven resonators for the encoding of bosonic codes.
As it has been recently proposed \cite{gravina2022critical,PRLLieu20}, detuning and critical phenomena may play a key role in storing quantum information. 
The metastability of the vacuum may also prove an obstacle to a rapid and reliable initialization of bosonic qubits.

This work paves the way for future intriguing research directions. 
Among them, we mention the study of the dynamical properties of these systems in connection with quantum trajectories approaches, and the emergence of chaotic behavior in highly nonlinear quantum resonators.

\section*{Acknowledgements}
We thank G. Rastelli and L. Gravina for the useful discussions. We acknowledge the help of A. Mercurio in the optimization of the numerical codes.
This work was supported by the Swiss National Science Foundation through Project No. 200020\_185015, and was conducted with the financial support of the EPFL Science Seed Fund 2021, PNRR MUR project PE0000023-NQSTI, Provincia Autonoma di Trento and 
from MUR under the PRIN2022 project 2022FLSPAJ (TANQU).
\appendix

\section{Interaction, nonlinearities, and $n$-photon drives in superconducting circuits}
\label{app:impl}

Let us consider a standard LC resonator characterized by the Hamiltonian:
\begin{equation}
\label{Eq:ham_Kerr}
    \hat{H}_{\rm cav}=\omega \hat{a}^\dagger \hat{a}, 
\end{equation}
where $\omega$ is the resonator frequency.

Non-quadratic (i.e., interaction) terms can emerge by considering the action of nonlinear elements.
For instance, nonlinearity can be obtained by quantizing the flux in Josephson junctions potentials of the form 
\begin{equation}\label{Eq:Expansion_cosine}
   E_J \cos\left(\frac{\phi}{\phi_0} \right) \simeq E_J \left( 1 - \frac{\phi^2 }{2 \phi_0 ^2}  + \frac{\phi^4}{24 \phi_0^4}  - \frac{\phi^6}{720 \phi_0^6}\right)+ \dots,
\end{equation}
where $\phi$ is the flux coordinate of the circuit at the junction and $\phi_0$ is the magnetic flux quantum.
In most implementations, the expansion in \eqref{Eq:Expansion_cosine} can be stopped at the $\phi^4$ order.
If the Josephson junction belongs to a single resonator, by substituting $\phi/\phi_0 \propto \hat{a}+\hat{a}^\dagger$, and discarding counter-rotating terms, which are out of resonance, one obtains the Kerr resonator Hamiltonian, reading  
\begin{equation}
\label{Eq:ham_Kerr_2}
    \hat{H}_{\rm Kerr}= \tilde{U}_1 \hat{a}^\dagger \hat{a} +\frac{U_2}{2} \left(\hat{a}^\dagger\right)^{2} \left(\hat{a}\right)^{2}.
\end{equation}

In $ n$-photon-driven systems, photons are coherently exchanged between the resonator and a set of external fields, $n$ at the time.
While single-photon drive (i.e., of the form $\hat{a}+\hat{a}^\dagger$) can emerge by, e.g., capacitive coupling an incoming wave-guide with the cavity, higher order drive requires to be mediated by nonlinear elements.
For instance, such $n$-drive terms can be derived from the expansion in \eqref{Eq:Expansion_cosine} if the Josephson junction is shared by several modes, so that $\phi= \sum_k \phi_k$, where $\phi_k$ represents the flux coordinate of each one of the modes.
For instance, two-photon drives can be achieved by standard four-wave mixing,
rewriting $\phi = \phi_a +\phi_b+ \phi_c$, where $\phi_a\propto \hat{a}+\hat{a}^\dagger$ is the field within the resonator, and $\phi_b \propto \hat{b} + \hat{b}^\dagger$ and $\phi_c \propto \hat{c} + \hat{c}^\dagger$ are auxiliary modes.
If the mode $b$ ($c$) is driven and evolves on a timescale much faster than the typical time scales of the $a$ mode, one can substitute the operators $\hat{b}$ ($\hat{c}$) with a c-number oscillating at the driving frequency $\omega_b$ ($\omega_c$) via an adiabatic elimination, reading $\hat{b} \to b e^{i \omega_b t}$ ($\hat{c} \to c e^{i \omega_c t}$ ).
All in all, discarding again out-of-resonance terms, the Hamiltonian for the resonator resulting from the fourth-order expansion of the potential $\cos(\phi)$ would result in a nonlinear Hamiltonian $\hat{H}_{\rm NL}$, reading
\begin{equation}
\label{Eq:ham_N}
    \hat{H}_{\rm NL}=\hat{H}_{\rm Kerr} + G_2 \left[\hat{a}^2  e^{2 i \omega_p t}+ \left(\hat{a}^\dagger\right)^2 e^{-2  i \omega_p t}\right].
\end{equation}
By passing in the frame rotating at the drive frequency and re-absorbing the contribution to the energy frequency in the term $U_1= \tilde{U}_1 -\omega_p=-\Delta$, the Hamiltonian finally reads
\begin{equation}
    \hat{H}_{n=2} = U_1 \hat{a}^\dagger \hat{a} + \frac{U_2}{2} \left( \hat{a}^\dagger \right)^2 \hat{a}^2 + G_2 \left[\hat{a}^2  + \left(\hat{a}^\dagger\right)^2 \right].
\end{equation}

Through similar procedures, high-order expansion of nonlinear terms ($k$-wave mixing with $k>n$), can result (in principle) in $n$-photon drives.
\textit{By including  such terms, one needs to include also the corresponding nonlinearities}. 
As detailed in the main text, such nonlinearities can play a fundamental role in determining the nature of the transition.

\section{An efficient algorithm for block-diagonalizing the Liouvillian in the presence of $Z_n$ symmetries}
\label{App:method}

We introduce here a simple algorithm to block diagonalize system displaying a $Z_n$ symmetry. Although we describe it for a weakly symmetric case, its extension to a strong symmetry is straightforward.

The Liouvillian admits an abstract definition of its spectrum via \eqref{Eq:Spectrum}. 
To numerically obtain the eigenvalues and eigenoperators one needs to explicit the matrix form of $\LL$.
For a finite-dimensional Hilbert space, one can construct such a matrix via
\begin{equation}\label{Eq:Liouvillian_matrix_form}
\begin{split}
    \LL &= -i\left(\hat{H} \otimes \hat{\mathds{1}}-\hat{\mathds{1}} \otimes \hat{H}^{\mathrm{T}}\right) \\ 
     & \quad + \sum_{j=1}^{3} \left( \hat{L}_j \otimes \hat{L}^{*}_j-\frac{\hat{L}_j^\dagger\hat{L}_j \otimes \hat{\mathds{1}}+ \hat{\mathds{1}} \otimes \hat{L}_j^{\rm T}\hat{L}_j^{*}}{2} \right), 
\end{split}
\end{equation}
where $\hat{L}_j^{\rm T}$ represents the transpose of $\hat{L}_j$.
The spectrum of the Liouvillian can then be directly obtained by diagonalizing the matrix representation of $\LL$.
For infinite dimensional spaces (i.e., those of bosonic systems), one needs to introduce a cutoff in the Hilbert space $N_c$.
That is, one projects the true infinite-dimensional Hamiltonian and jump operators onto the space spanned by the Fock states $\ket{n}$ for $n \in [0, N_c)$, and assumes that the matrix elements of any operator for $n \in [N_c, \infty)$ are zero.

Since $[\mathcal{Z}_{n},\LL]=0$, all the $\eig{i}$ are eigenoperators of $\mathcal{Z}_{N}$. And since $\mathcal{Z}_{n}$ admits $n$ different eigenvalues,  it is always possible to block-diagonalize the Liouvillian into (at least) $n$ smaller blocks.
Each block $\mathcal{L}_n$ describes completely the physics of each symmetry sector of the full Liouvillian $\LL$.
Normally, to put the Liouvillian in its block-diagonal form one would construct the basis of a symmetry sector determining the eigenoperators $\hat{\zeta}_i$ of $\mathcal{Z}_n$ and project the Liouvillian onto the correct symmetry sector obtaining the matrix elements
\begin{equation}\label{Eq:symmetry_sectors_slow}
    \LL_{i, j} = \operatorname{Tr}\left[\hat{\zeta}_i^\dagger \left(\LL \hat{\zeta}_j \right)\right].
\end{equation}
Even if, in principle, correct, this process is extremely slow and inefficient since the Liouvillian is a very sparse and large matrix.

Instead of applying \eqref{Eq:symmetry_sectors_slow}, we notice that the Fock basis is already the basis of eigenstates of $\mathcal{Z}_n$, as it follows from \eqref{Eq:condition_symmetry}. That is, when using \eqref{Eq:Liouvillian_matrix_form}, we are using the correct basis to obtain the block-diagonal form of the Liouvillian, simply we are considering the basis in the wrong order.
Hence, the Liouvillian is a permutation of rows and columns away from being block diagonal, and the algorithm that we seek is one that efficiently finds the correct permutation matrix $\mathcal{P}$ which transforms  $\LL$ into its block diagonal form, whenever such a transformation is possible.

The main idea is to model the block diagonalization problem as an equivalent graph-theoretic problem.
\begin{enumerate}
    \item $\mathcal{L}$ is written as the adjacency matrix of an undirected graph;
    \item Each block in the block diagonal form is a single connected component in the graph; thus, the problem boils down to finding each connected component in the graph.
    \item We then use the Breadth First/Depth First search algorithm 
    consecutively to obtain the permutation matrices and the indices of the blocks. 
    The time to perform this task (i.e., its computational complexity) is linear in the number of nodes in the graph.
    \item We use the permutation matrix to produce each block $\LL_{i}$ such that ${\LL=\mathcal{P}\operatorname{diag}\{\LL_{1}\dots \LL_{n} \}\mathcal{P}^{\rm T}}$.
\end{enumerate}
The key factor in the numerical speedup comes from the fact that obtaining the permutation matrix $\mathcal{P}$ requires a number of operations linear in the number of nonzero elements of the Liouvillian, which is a very sparse matrix [c.f. \eqref{Eq:Liouvillian_matrix_form}].


%

\end{document}